\documentclass[11pt,dvipsnames]{article}

\pdfoutput=1
\usepackage[utf8]{inputenc}
\usepackage[linktocpage,bookmarks,plainpages=false,pdfborder={0 0 0},pdfborderstyle={}]{hyperref}

\DeclareMathAlphabet{\mathpzc}{OT1}{pzc}{m}{it}
\usepackage{makecell} 
\usepackage{graphicx}
\usepackage{amssymb}
\usepackage{textcomp}
\usepackage{amsmath}
\usepackage{cite}
\usepackage{geometry}                
\usepackage{graphicx}
\usepackage{amssymb}
\usepackage{epstopdf}
\usepackage{pgfplots}
\usepackage{tikz-3dplot}

\newcommand{\Rep}[1]{\ensuremath{\boldsymbol{\underline{#1}}}}
 
\tdplotsetmaincoords{60}{115}
\pgfplotsset{compat=newest}
\usepackage{mathrsfs}
\global\long\def\Ex#1{{E}_{#1}}%

\makeatletter
\renewcommand*\env@matrix[1][\arraystretch]{%
  \edef\arraystretch{#1}%
  \hskip -\arraycolsep
  \let\@ifnextchar\new@ifnextchar
  \array{*\c@MaxMatrixCols c}}
\makeatother
\usepackage{booktabs}

\usepackage{geometry}                
\usepackage{graphicx}
\usepackage{amssymb}
\usepackage{epstopdf}
\usepackage{amsmath}

\usepackage{bm}
\usepackage{times}
\usepackage{epsfig}
\usepackage{color}
\usepackage{array,multirow}

\newcommand{\trix}[1]{\left(\begin{array}{#1}}
\newcommand{\notrix}{\end{array}\right)}
\newcommand{\comment}[1]{}

\def\beq{\begin{equation}}
\def\eeq{\end{equation}}
\def\bea{\begin{eqnarray}}
\def\eea{\end{eqnarray}}

\usepackage[normalem]{ulem}

\usepackage[T1]{fontenc}

\usepackage[utf8]{inputenc}

\usepackage{mathtools} 

\usepackage{amsfonts}

\usepackage{ amssymb }

\usepackage[toc,page,title]{appendix}

\usepackage{xcolor}

\usepackage[compat=1.1.0]{tikz-feynman}

\usepackage{graphicx}

\usepackage{subcaption}

\usepackage{float}

\usepackage{caption}

\usepackage{slashed}

\usepackage{breqn}

\usepackage{amsmath}
\usetikzlibrary{arrows.meta,angles,quotes,calc}

\global\long\def\Z{\mathbb{Z}}%

\usepackage{subcaption}
\usepackage{float}
\usepackage{caption}
\allowdisplaybreaks

\usepackage[utf8]{inputenc}

\usepackage{mathtools} 

\usepackage{amsfonts}

\usepackage{ amssymb }

\usepackage[toc,page,title]{appendix}

\usepackage{xcolor}

\usepackage[compat=1.1.0]{tikz-feynman}

\usepackage{graphicx}

\usepackage{subcaption}

\usepackage{float}

\usepackage{caption}

\numberwithin{equation}{section}

\usepackage{slashed}
\begin{document}

\begin{titlepage}
\title{\LARGE {FIMP Dark Matter in Heterotic M-Theory}\\[.3cm]}
                       
\author{{
   Sebastian Dumitru
   and Burt A.\,Ovrut}\\[0.8cm]
   {${}$\it Department of Physics, University of Pennsylvania} \\[.1cm]
   {\it Philadelphia, PA 19104, USA}
 }      

\date{}

\maketitle

\begin{abstract}
\noindent
Within the context of $N=1$ supersymmetric heterotic $M$-theory, we present a ``freeze-in'' mechanism for producing dark matter via a ``moduli portal'' between the observable and hidden sectors. It is assumed that the observable sector consists of the MSSM or some physically acceptable extension of it, while the hidden sector is chosen to satisfy all physical and mathematical constraints. Dark matter production processes are examined for two fundamental types of hidden sectors; those whose gauge bundle structure group contains an anomalous $U(1)$ and those whose structure group is non-Abelian and anomaly free. The couplings of the dilaton and the ``universal'' modulus to all fields of the observable and hidden sectors are presented and analyzed. These interactions are then combined to produce a moduli portal from a thermal bath of observable sector particles to the hidden sector. 
These processes are then analyzed for both anomalous and non-anomalous cases. It is shown that only the uncharged hidden sector matter scalars can play the role of dark matter and that these are predominantly produced during the ``reheating'' epoch on the observable sector. Within the context of both an anomalous and non-anomalous hidden sector, we calculated the dark matter ``relic density''. We show that in both cases, for a wide choice of moduli vacua, one can correctly predict the observed relic density. For the anomalous $U(1)$ case, we choose a specific physically acceptable vacuum within the context of the $B-L$ MSSM and show that one precisely obtains the measured dark matter relic abundance.

\noindent

\let\thefootnote\relax\footnotetext{\noindent sdumitru@sas.upenn.edu, ovrut@elcapitan.hep.upenn.edu}

\end{abstract}

\thispagestyle{empty}
\end{titlepage}

\tableofcontents

\section{Introduction}

We present a scenario in which the observed Dark Matter (DM) consists of Feebly Interacting Massive Particles (FIMPs), produced non-thermally by a so-called "freeze-in" mechanism. In contrast to the usual freeze-out scenario, frozen-in FIMP DM interacts very weakly with the particles in the visible sector and, therefore, it never attains thermal equilibrium with the visible baryon-photon fluid in the early Universe. FIMP DM models gained popularity recently, following the null results of WIMP DM searches~\cite{Bernal:2017kxu, Hall:2009bx} (for a review of WIMP searches and models, see \cite{Arcadi:2017kky}).

Heterotic M-theory is an appealing context to analyze this DM production mechanism because the structure of its 11-dimensional vacuum is that of a double domain wall~\cite{Lukas:1998yy}. Following the compactification of the extra dimensions, all matter fields of the Standard Model, including its supersymmetric extensions, are found on the so-called  "observable" wall.  On the "hidden" wall, on the other hand, we find an analog spectrum, consisting of chiral matter fields, as well as vector supermultiplets. The existence of such a non-trivial hidden sector is necessary to cancel all anomalies in the heterotic vacuum with preserved supersymmetry~\cite{Lukas:1999nh}. These "hidden fields" have no gauge interactions in common with the visible sector fields in the 4D low-energy theory; their only possible interactions with the SM are gravitational, making them ideal DM candidates. Furthermore, considering the weak nature of their interaction with the visible sector, it is only possible for such "hidden dark matter" to be produced by a "freeze-in" mechanism, via moduli portals of the type analyzed in \cite{Chowdhury:2018tzw,Dutra:2019nhh}. In the following, we will review some specific constructions of the M-theory heterotic vacua, which successfully reproduced the observed spectrum of particles.

Heterotic $M$-theory is eleven-dimensional Horava-Witten theory \cite{Horava:1995qa,Horava:1996ma} dimensionally reduced to five-dimensions by compactifying on a Calabi-Yau (CY) complex threefold. It was first introduced in \cite{Lukas:1997fg} and discussed in detail in \cite{Lukas:1998yy, Lukas:1998tt}. Five-dimensional heterotic $M$-theory consists of two four-dimensional orbifold planes separated by a finite fifth-dimension. The two orbifold planes, each with an $E_{8}$ gauge group, are called the observable and hidden sectors respectively~\cite{Lukas:1997fg, Lukas:1998yy, Lukas:1998tt, Donagi:1998xe,Ovrut:2000bi}. By choosing a suitable CY threefold, as well as an
appropriate holomorphic vector bundle~\cite{Donagi:1999gc} on the CY compactification at the observable sector, one can find realistic low energy $N=1$ supersymmetric particle physics models. A number of such realistic theories on the observable sector have been constructed. See, for example~\cite{Braun:2005nv,Braun:2005bw,Braun:2005ux,Bouchard:2005ag,Anderson:2009mh,Braun:2011ni,Anderson:2011ns,Anderson:2012yf,Anderson:2013xka,Nibbelink:2015ixa,Nibbelink:2015vha,Braun:2006ae,Blaszczyk:2010db,Andreas:1999ty,Curio:2004pf}.

In particular, in \cite{Braun:2005bw,Braun:2005zv,Braun:2005nv, Braun:2006ae} it was shown that one can obtain the exact MSSM spectrum if one compactifies Horava-Witten theory on a specific Schoen Calabi--Yau threefold $X$, with a particular holomorphic gauge bundle with structure group  $SU(4)\subset E_8$ on the observable sector. This $SU(4)$ bundle breaks the $E_8$ group down to 
\begin{equation}
  E_8 \to Spin(10) \ .
  \label{13}
\end{equation}
Hence, $Spin(10)$ is the ``grand unified'' group of the observable sector. This GUT group is then further broken at scale $\langle M_{U}\rangle=3.15 \times 10^{16}$ GeV to the low energy gauge group 
by turning on two flat Wilson lines, each associated with a different $\Z_3$ factor of the $\Z_3 \times \Z_3$ holonomy of $X$. Doing this preserves $N=1$ supersymmetry in the four-dimensional effective theory, but breaks the observable sector gauge group down to
\begin{equation}
  Spin(10) 
  \to 
  SU(3)_C \times SU(2)_L \times U(1)_Y \times U(1)_{B-L} \ .
  \label{17}
\end{equation}
That is, the gauge group 
is that of the MSSM with an addition $U(1)_{B-L}$ gauge factor. The spectrum was explicitly computed and extensively studied in \cite{Braun:2005nv,Braun:2005bw}. It is found to contain exactly the three quark and lepton families of the  MSSM--including three additional right-handed neutrino chiral multiplets, one per family--as well as the conventional $H_{u}$ and $H_{d}$ Higgs doublet supermultiplets. There are no exotic fields or vector-like pairs. This phenomenologically realistic vacuum is called the $B-L$ MSSM. We note that this model was also constructed from a ``bottom-up'' perspective, using phenomenological, low energy considerations, in~\cite{Ambroso:2009jd,Marshall:2014kea,Marshall:2014cwa,Ovrut:2012wg,Ovrut:2014rba,Barger:2008wn,FileviezPerez:2009gr}.

We want to emphasize that the ``freeze-in'' mechanism of producing dark matter discussed in this paper is, in principle, {\it valid for any realistic heterotic $M$-theory vacuum}. Indeed, the formalism we develop here will be given in generic terms and is generally applicable. However, once the formalism is presented, it will be useful to apply it to a specific vacuum to explicitly calculate cosmological quantities--such as the dark matter relic abundance. The specific vacuum we will use is the $B-L$ MSSM. We choose this theory for two reasons. First, as discussed above, the observable sector is phenomenologically realistic--not only having the correct low energy spectrum, but also giving the appropriate masses of the electroweak gauge bosons, the Higgs mass, and so on. Secondly, the $B-L$ MSSM was shown in \cite{Deen:2016zfr,Cai:2018ljy} to admit a viable inflation model, with a linear combination of the neutral Higgs field and the third family right-handed and left-handed sneutrinos as the inflaton. Hence, obtaining suitable dark matter production in this context would be compatible with a realistic cosmological theory in the observable sector. A brief outline of this Higgs-Sneutrino inflation theory--along with a discussion of several of its properties relevant to this paper--is presented in Appendix A.

A generic heterotic $M$-theory vacuum will, by construction, contain both a dilaton and several geometric moduli chiral superfields. The geometric moduli correspond to both the K\"ahler and complex structure moduli of the Calabi-Yau threefold on which the Horava-Witten theory was compactified. In a series
of papers~\cite{Anderson:2010mh, Anderson:2011cza,Anderson:2011ty}, it was shown, for a range of examples, that
one can fix all complex structure moduli masses to be of the order of the compactification scale. A more general discussion was presented in~\cite{Cicoli:2013rwa}. In this paper, based on these results, we will assume that the masses of all complex structure moduli are near the compactification scale and, therefore, can be neglected in our analysis. However, both the dilaton and the K\"ahler moduli fields can have lighter masses--on the order of the $N=1$ supersymmetry breaking scale--and will play a fundamental role in this paper. In the context of $N=1$ supergravity, these dilaton/moduli superfields couple to both the vector and chiral matter superfields on the observable sector--which we will assume form the MSSM or a realistic extension thereof--as well as interact with the gauge and matter supermultiplets on the hidden sector. That is, the dilaton and K\"ahler moduli chiral supermultiplets create a ``portal'' which connects the observable and hidden sectors. Specifically, vector and matter multiplets on the observable sector can decay to dilaton/moduli superfields which, after traveling across the finite fifth dimension of the heterotic vacuum, in turn, decay into vector and matter superfields of the hidden sector. It is the purpose of this paper to show that a subset of the hidden sector supermultiplets can play the role of cosmological dark matter and, importantly, that these explicit dilaton/moduli portal interactions can exactly produce the observed value for the ``relic dark matter abundance''.

Specifically, we will do the following. In Section 2, we present the gauge and matter superfields of the observable and hidden sectors, as well as the perturbative superpotentials associated with each sector. This is followed by a brief discussion of the dilaton $S$ and the K\"ahler chiral supermultiplets in the generic case of a CY threefold with $h^{1,1} \geq 1$. However, to be consistent with the inflationary cosmology model in \cite{Deen:2016zfr,Cai:2018ljy}--which we refer to throughout this paper-- we then restrict our theory to the single ``universal'' modulus $T$. We introduce the non-perturbative superpotential for $S$ and $T$ that can be generated by various mechanisms, such as gaugino condensation. Finally, we give the K\"ahler potentials and various gauge couplings that arise in the heterotic $M$-theory vacuum. In Section 3, we give a detailed discussion of the $S$ and $T$ moduli as ``portals'' between the observable and hidden sectors. We begin by presenting the explicit $N=1$ supergravity couplings of both observable and hidden sector gauge and chiral supermultiplets to the $S$ and $T$ moduli. We then compute the moduli mass eigenstates--first, for the case of an anomalous $U(1)$ gauge group in the hidden sector and, second, for a hidden sector without an anomalous $U(1)$. For each case, we present the explicit couplings of both observable and hidden sector gauge fields, gauginos and scalar and fermion matter to the moduli mass eigenstates. Using the results of Section 3, in Section 4 we combine these interactions and discuss the complete mechanisms for creating hidden dark matter via the moduli eigenstate portals. Again, this is carried out for the anomalous and non-anomalous hidden sector gauge groups respectively. In the anomalous case, it is shown that significant amounts of dark matter can only be created during the ``reheating epoch'' from the thermal bath of observable sector particles. Furthermore, the dominant mechanism produces scalar dark matter particles with a mass of order $10^{13}$ GeV. This corresponds to the so-called ``freeze-in'' mechanism. Similarly, in the non-anomalous case, dark matter production occurs during the reheating phase and, again, through the ``freeze-in'' mechanism predominantly produces scalar dark matter particles with a mass of order $10^{13}$ GeV. We end Section 4 by calculating the dark matter ``relic density'' for the anomalous and non-anomalous hidden sector cases respectively. We find that, in both cases, a reasonable choice of the dilaton and universal modulus vacua leads to the observed relic abundance.

Our analysis is, in principle, compatible with any $N=1$ supersymmetric observable sector theory of inflation.  However, in this paper--for specificity--we will assume that the cosmology on the observable sector is given by the Higgs-Sneutrino theory presented in \cite{Deen:2016zfr,Cai:2018ljy}. A brief outline of the relevant parts of that theory is presented in Appendix A. As mentioned above, it is necessary to find the moduli masses eigenstates for both the anomalous and non-anomalous hidden sectors. To help carry out this procedure in the more complicated non-anomalous case,
we present the necessary details in Appendix B. Finally, we give an explicit physical example of the dark matter production mechanism in the anomalous $U(1)$ case. We do this within the context of the $B-L$ MSSM heterotic theory. The relevant details of the $B-L$ MSSM theory are presented in Appendix C.

\section{Generic 4D Effective Theory}\label{sec:Effective theory}

As discussed above, a generic heterotic $M$-theory vacuum consists of two four-dimensional domain walls--the observable and hidden sectors respectively-- separated by a finite fifth dimension. Let us first consider the observable sector.

Below the unification scale, the observable sector consists of an $N=1$ supersymmetric theory with both gauge vector superfields as well as chiral matter superfields. In this paper, we will assume that this theory is phenomenologically realistic--that is, that both the gauge and matter superfields are those of the MSSM, or a realistic extension thereof. We will denote the vector superfields as $V_{(o)}^{a}=( A_{(o )\mu}^{a}, \Lambda_{(o)}^{a} )$, where the index $a$ runs over the adjoint representation of $SU(3)_{C} \times SU(2)_{L} \times U(1)_{Y}$ and any further extension of the standard model gauge group--such as $U(1)_{B-L}$ in the $B-L$ MSSM. The chiral matter superfields will be generically denoted by $\tilde{C}_{(o)}^{\cal{I}}=(C_{(o)}^{\cal{I}}, \Psi_{(o)}^{\cal{I}})$, where ${\cal{I}}=1, \dots,{\cal{N}}_{(o)}$. Here, ${\cal{N}}_{(o)}$ is the number of quark, lepton, and Higgs superfields of the MSSM,  plus any further multiplets introduced into the observable sector--such as the three right-handed sneutrino chiral multiplets of the $B-L$ MSSM. Finally, the generic form of the observable sector superpotential is given by
\begin{equation}
\label{burt1}
\tilde{W}^{\text{(obs)}}=\mu_{\cal{I} \cal{J}} \tilde{C}_{(o)}^{\cal{I}}\tilde{C}_{(o)}^{\cal{J}}+Y_{\cal{I}\cal{J}\cal{K}}\tilde{C}_{(o)}^{\cal{I}}\tilde{C}_{(o)}^{\cal{J}}\tilde{C}_{(o)}^{\cal{K}}\ 
\end{equation}
where in the MSSM $\mu_{\cal{I} \cal{J}}$ corresponds to the Higgs $\mu$-term and $Y_{\cal{I}\cal{J}\cal{K}}$ are the Yukawa couplings.

The hidden sector below the unification scale contains an $N=1$ supersymmetric theory whose precise form depends on the explicit choice of the CY threefold, the observable sector vector bundle and any five-branes in the fifth-dimensional interval. It is determined, given this input, by solving several explicit constraints--such as anomaly cancelation, the positivity of squared gauge couplings, and so on \cite{Ashmore:2020ocb}. This has been carried out in detail in several realistic heterotic $M$-theory models, including the $B-L$ MSSM theory.
The hidden sector vector bundle has structure group $G^{(2)}$, which breaks $E_8$ to the commutant subgroup $H^{(2)}$ in 4D.
In this section, we will consider a general scenario in which the low energy subgroup has the form $H^{(2)}= \mathcal{H}_2\times \mathcal{H}_1$. We do this for the following reason. Generically, we want the low energy group to contain a factor--denoted by $\mathcal{H}_2$--which can become strongly coupled, thus leading to gaugino condensation. The condensate induces non-zero F-terms in the 4D
effective theory which break SUSY globally~\cite{Choi:1997cm,Kaplunovsky:1993rd,Horava:1996vs,Lukas:1997rb,Nilles:1998sx,Binetruy:1996xja,Antoniadis:1997xk,Minasian:2017eur,Gray:2007qy,Lukas:1999kt,Font:1990nt}. On the other hand, we want the the low energy group to contain a sector--denoted by $\mathcal{H}_1$--which does not become strongly coupled. Note that 
if $G^{(2)}$  contains an anomalous $U(1)$ factor as in \cite{Ashmore:2020ocb, Dine:1987xk,Dine:1987gj,Anastasopoulos:2006cz,Blumenhagen:2005ga}, since $U(1)$ commutes with itself, $U(1) \subseteq \mathcal{H}_1$.  
We will denote the hidden sector vector superfields as $V_{(h)}^{a^{\prime}}=( A_{(h )\mu}^{a^{\prime}}, \Lambda_{(h)}^{a^{\prime}} )$, where the index $a^{\prime}$ runs over the adjoint representation of $\mathcal{H}_2\times \mathcal{H}_1$. The low energy chiral matter spectrum also depends on the explicit choice of the heterotic  $M$-theory model. In this paper, we will assume that the low energy chiral matter spectrum contains ${{N}}_{(h)}$ matter fields which are {\it singlets} under $\mathcal{H}_2$. We denote these $\mathcal{H}_2$ singlet matter chiral multiplets as $\tilde{C}_{(h)}^{L}=(C_{(h)}^{L}, \Psi_{(h)}^{L}), L=1,\cdots,{{N}}_{(h)}$.
In our analysis, any chiral matter multiplets transforming non-trivially under the $\mathcal{H}_2$ subgroup do not play any role and will, henceforth, be ignored. Finally, the hidden sector superpotential is of the generic form
\begin{equation}
\label{burt2}
W^{\text{(hid)}}= m_{LM}\tilde{C}_{(h)}^{L}\tilde{C}_{(h)}^{M}+\lambda_{KLM}\tilde{C}_{(h)}^K\tilde{C}_{(h)}^L\tilde{C}_{(H)}^M  \ .
\end{equation}

In addition to the matter and gauge superfields on the observable and the hidden sectors, the $D=4$ low energy effective theory contains 1) a universal dilaton chiral supermultiplet $\tilde{S}=(S, \Psi_{S})$ and 2) the geometric moduli superfields--that is, the complex structure and K\"ahler moduli--associated with the specific choice of CY threefold. We assume that the complex structure fields have been already stabilized with masses near the unification scale. Hence, we will ignore them. However, the K\"ahler moduli, as we discuss below, can be much lighter and play a crucial role in our analysis as the moduli portal between the observable and hidden sectors. The number of K\"ahler moduli is given by the dimension of the $H^{1,1}$ cohomology of the CY. For different realistic heterotic $M$-theory vacua, the value of $h^{1,1}$ can vary, typically with $ h^{1,1}>1$. For example, the Schoen threefold of the $B-L$ MSSM has $h^{1,1}=3$. One could analyze dark matter production using all $T^{i}$, $i=1,\dots,h^{1,1}$ K\"ahler moduli. This will be carried out elsewhere. Here, however, we will simplify the formalism by using a result given in \cite{Lukas:1998tt}. There it was shown that an $h^{1,1}>1$ heterotic $M$-theory vacuum is consistent with a vacuum defined by an $h^{1,1}=1$ ``universal'' modulus $T$. Loosely speaking, this corresponds to fixing the values of all K\"ahler moduli up to a universal ``breathing'' mode. Since the inflationary cosmology presented in \cite{Deen:2016zfr,Cai:2018ljy} is written in terms of this universal modulus, we will do the same in this paper. That is, in addition to the dilaton chiral multiplet $\tilde{S}$, we will consider a universal K\"ahler chiral multiplet $\tilde{T}=(T, \Psi_{T})$.
At the perturbative level there is no superpotential associated with the $S$ and $T$ moduli. However, as we discuss below, non-perturbative effects, such as gaugino condensation on the hidden sector or five-brane instantons, can produce a non-vanishing moduli superpotential
\begin{equation}
\label{burt3}
W^{\text{(mod)}}={\hat{W}}(S,T) \ .
\end{equation}

Finally, we note that all of the above superfields couple to $N=1$ supergravity and, hence, to the graviton, gravitino and the gravitational auxiliary fields M and $b_{a}$. However, these couplings are strongly suppressed relative to the moduli interactions and, hence, we will ignore them in this paper. An important implication of this is that the contribution of the auxiliary field M to the scalar potential and various couplings \cite{Ovrut:1981zu} is very small and does not contribute to the dark matter production mechanism discussed in this paper.

The K\"ahler potentials for the heterotic $M$-theory vacua presented above can be completely determined using the formalism and definitions presented in~\cite{Lukas:1997fg,Lukas:1998tt} (see also~\cite{Brandle:2003uya} for the an explicit $h^{1,1}=3$ case). 
Specifically, one finds that to order $\kappa^{2/3}_4$ the K\"ahler potential takes the form
\begin{align}
K=-\kappa_4^{-2}\ln(S+\bar S)-3\kappa_4^{-2}\ln\left(T+\bar T-\kappa_{4}^{2} \mathcal{G}_{{\cal{I}}{\bar{\cal{J}}} } C_{(o)}^{\cal{I}}\bar{C}_{(o)}^{\bar {\cal{J}}}-\kappa_{4}^{2}\mathcal{G}_{{{L}}{\bar{M}} } C_{(h)}^{L}\bar{C} \ _{(h)}^{\bar {M}}\right) \label{bl1} \ ,
\end{align}
where $\mathcal{G}_{{\cal{I}}{\bar{\cal{J}}}}$ and $\mathcal{G}_{{L}{\bar{M}}}$ are dimensionless hermitian matrices which generically depend on the moduli. The dilaton $S$ and the universal modulus $T$ are chosen to be dimensionless, whereas the scalar fields $C_{(o)}^{\cal{I}}$ and $C_{(h)}^{L}$ have mass dimension one.
Expanding to linear order in the matter field terms, we can write the K\"ahler potential in the form
\begin{equation}
\label{burt5}
K=K_S+K_T+	Z_{{\cal{I}}{\bar{\cal{J}}} } C_{(o)}^{\cal{I}}\bar{C}_{(o)}^{\bar {\cal{J}}}+Z_{{{L}}{\bar{M}} } C_{(h)}^{L}\bar{C}_{(h)}^{\bar {M}}\ ,
\end{equation}
where we define $K_S$ and $K_T$ to be the pure moduli parts of the K\"ahler potential
\begin{equation}
\label{burt6}
\begin{split}
K_S&=-\kappa_4^{-2}\ln(S+\bar S)\ ,\\
K_T&=-3\kappa_4^{-2}\ln \left(T+\bar T\right)
\end{split}
\end{equation}
respectively, and
\begin{equation}
\label{burt7}
\begin{split}
Z_{\cal{I}\bar{\cal{J}}}&=3e^{\kappa_4^2K_T/3}\mathcal{G}_{\cal{I}\bar{\cal{J}}} \ ,\\
Z_{L \bar{M}}&=3e^{\kappa_4^2K_T/3}\mathcal{G}_{L\bar{M} }\ 
\end{split}
\end{equation}
are the moduli-dependent metrics. The hermitian matrices $Z_{\cal{I}\bar{\cal{J}}}$ and $Z_{L \bar{M}}$ can always be chosen to be diagonal with real coefficients.
Note , however, that when computing explicit amplitudes later in this paper, we will assume, for simplicity, that the
matter field internal metrics have the diagonal form $ \mathcal{G}_{{\cal{I}}{\bar{\cal{J}}} }=\delta_{{\cal{I}}{\bar{\cal{J}}} }$, $\mathcal{G}_{L{\bar{M}}}=\delta_{L{\bar{M}}}$. Modifying this to allow diagonal matrices with non-unit entries, for example, does not alter the conclusions of this paper. See, for example, the discussion at the end of subsection 4.4 and Appendix C. Finally, to order $\kappa^{2/3}_4$ gauge threshold corrections~\cite{Lukas:1997fg}, the gauge kinetic functions on the observable and the hidden sectors are given by
\begin{equation}
\label{burt8}
f_1=f_2=S\ .
\end{equation}
 These gauge kinetic functions determine the values of the unified gauge couplings $g_{1}$ and $g_{2}$ on the observable and hidden sector respectively at the unification scale $M_U=3.15\times 10^{16}$GeV. That is,
\begin{equation}
{g_{1}^{2}}=\frac{\pi \hat{\alpha}_{\text{GUT}}}{\text{Re} f_{1}} , \qquad  {g_{2}^{2}}=\frac{ \pi \hat{\alpha}_{\text{GUT}}}{\text{Re} f_{2}}  \ ,
\label{umb1}
\end{equation}
where $\hat \alpha_{\text{GUT}}$ is a moduli-dependent parameter. 

\section{Moduli Portals}\label{sec:modPortal}

In this section, we identify the ``moduli portals'' that couple the observable sector to the hidden sector. These couplings are relevant in the process of producing DM via the freeze-in mechanism, as will be explained in the next section. Such moduli field portals were also analyzed in \cite{Chowdhury:2018tzw,Dutra:2019nhh}, although not in a specific physical context. These couplings will be expressed in terms of 4-components Dirac spinors,
\begin{equation}
\psi_{(o)}^{\mathcal I}=\begin{pmatrix} \Psi_{(o)}^{\mathcal I}\\ \bar \Psi_{(o)}^{\mathcal I\dag} \end{pmatrix}\ ,\quad
\psi_{(h)}^{L}=\begin{pmatrix} \Psi_{(h)}^{L}\\ \bar \Psi_{(h)}^{L\dag} \end{pmatrix}\ ,\quad 
\lambda^a_{(o)}=\begin{pmatrix} \Lambda^a_{(o)}\\  \Lambda^{a\dag}_{(o)}\end{pmatrix}\ ,\quad 
\lambda^a_{(h)}=\begin{pmatrix} \Lambda^{a^\prime}_{(h)}\\   \Lambda^{a^\prime\dag}_{(h)}\end{pmatrix}\ .
\end{equation}

\subsection{ Matter Coupling to Moduli}

We begin our analysis by considering the 4D supergravity Lagrangian. The 4D effective theory contains the following kinetic terms for the matter scalars
\begin{equation}
\begin{split}
\label{eq:kinetic_terms_infl}
\mathcal{L}&\supset  -3e^{\kappa_4^2K_T/3}\mathcal{G}_{\cal I\bar {\cal J}} \partial_\mu C_{(o)}^{\cal I}\partial^\mu \bar C_{(o)}^{\bar {\cal J}}-
3e^{\kappa_4^2K_T/3}\mathcal{G}_{L\bar {M}} \partial_\mu C_{(h)}^{L}\partial^\mu \bar C_{(h)}^{\bar {M}}\ ,
\end{split}
\end{equation}
and the matter fermions
\begin{equation}
\begin{split}
\mathcal{L}&\supset  -3i e^{\kappa_4^2K_T/3}\mathcal{G}_{\cal I\bar {\cal J}}  \psi_{(o)}^{\bar{ \cal J}\dag}\slashed{\partial} \psi_{(o)}^{\cal I}
-3i e^{\kappa_4^2K_T/3}\mathcal{G}_{L\bar {M}}  \psi_{(h)}^{\bar{ M}\dag}\slashed{\partial} \psi_{(h)}^{L}
\end{split}
\end{equation}
of both the observable and hidden sectors.
The complex scalar components of the moduli superfields decompose as
\begin{equation}
\begin{split}
\label{eq:def_scalar_intro}
& S=s+i\sigma\ ,\\
&T=t+2i\chi\ ,\\
\end{split}
\end{equation}
where $\sigma$ and $\chi$ are the dilaton axion and the universal K\"ahler axion respectively. It follows from \eqref{burt5}-\eqref{burt7} that both the observable and hidden sector matter fields couple to $t^{-1}$. As discussed in Appendix A, non-perturbative effects, such as gaugino condensation, can stabilize the $s$ and $t$ moduli at fixed VEVs $\langle s \rangle$ and $\langle t \rangle$ respectively. After expanding $t$ to linear order around its VEV, that is, $t\rightarrow \langle t\rangle +\delta t$, and performing the field redefinitions $C_{(o)}^{\cal I}\rightarrow
\sqrt{\frac{3}{\langle T+\bar T\rangle}}C_{(o)}^{\cal I}$, $C_{(h)}^{L}\rightarrow
\sqrt{\frac{3}{\langle T+\bar T\rangle }}C_{(h)}^{L}$, as well as $\psi_{(o)}^{\cal I}\rightarrow
\sqrt{\frac{3}{\langle T+\bar T\rangle}}\psi_{(o)}^{\cal I}$, $\psi_{(h)}^{L}\rightarrow
\sqrt{\frac{3}{\langle T+\bar T\rangle }}\psi_{(h)}^{L}$, we obtain
\begin{equation}
\begin{split}
\label{eq:int_sc_1}
\mathcal{L}&\supset  -\mathcal{G}_{{\cal I}\bar {\cal J}}  \partial_\mu C_{(o)}^{\cal I}\partial^\mu \bar C_{(o)}^{\bar {\cal J}}
-\mathcal{G}_{L\bar {M}} \partial_\mu C_{(h)}^{L}\partial^\mu \bar C_{(h)}^{\bar {M}}\\
 &+\frac{ 1}{\langle t\rangle }\delta t\mathcal{G}_{{\cal I}\bar {\cal J}}\partial_\mu C_{(o)}^{\cal I}\partial^\mu \bar C_{(o)}^{\bar {\cal J}}
 +\frac{ 1}{\langle t\rangle }\delta t \mathcal{G}_{L\bar {M}} \partial_\mu C_{(h)}^{L}\partial^\mu \bar C_{(h)}^{\bar {M}}
\end{split}
\end{equation}
and
\begin{equation}
\begin{split}
\label{eq:int_sc_2}
\mathcal{L}&\supset  -i\mathcal{G}_{{\cal I}\bar {\cal J}}  \psi_{(o)}^{\bar {\cal J}\dag}\slashed{\partial} \psi_{(o)}^{{\cal I}}- i\mathcal{G}_{L\bar M}  \psi_{(h)}^{\bar {M}\dag}\slashed{\partial} \psi_{(h)}^{{L}}
\ \\
&+i\frac{ 1}{\langle t\rangle }\delta t\mathcal{G}_{{\cal I}\bar {\cal J}} \psi_{(o)}^{\bar {\cal J}\dag}\slashed{\partial} \psi_{(o)}^{I}+i\frac{ 1}{\langle t\rangle }\delta t\mathcal{G}_{L\bar {M}}  \psi_{(h)}^{\bar {M}\dag}\slashed{\partial} \psi_{(h)}^{L}\ .
\end{split}
\end{equation}
These terms source the couplings of the matter scalars and fermions to the $t$ modulus perturbation. Note that the kinetic terms do not source couplings to either the axionic component of the $T$ modulus or to either component of the dilaton $S$.

\subsection{Gauge Multiplets Coupling to Moduli}

In the previous subsection, we wrote down the relevant couplings of scalar and fermionic matter to the moduli fields--both in the observable and the hidden sectors. In this subsection, we will compute the couplings of the supersymmetric gauge fields, that is, both the gauge connection and the associated gauginos, to the moduli fields.
The 4D effective Lagrangian has the gauge kinetic couplings
\begin{equation}
\begin{split}
\mathcal{L}\supset& -\frac{1}{4\pi\hat \alpha_{\text{GUT}}}\text{Re}(f_1)F_{(o)}^{\mu \nu\> a}F^a_{\mu \nu\>(o)} +\frac{i}{4\pi\hat \alpha_{\text{GUT}}}\text{Im}(f_1)F_{(o)}^{\mu \nu\> a}\tilde F^a_{\mu \nu\>(o)}\\
&-\frac{1}{4\pi\hat \alpha_{\text{GUT}}}\text{Re}(f_2)F_{(h)}^{\mu \nu\> a^{\prime}}F^{a^{\prime}}_{\mu \nu\>(h)}
+\frac{i}{4\pi\hat \alpha_{\text{GUT}}}\text{Im}(f_2)F_{(h)}^{\mu \nu\> a^{\prime}}\tilde F^{a^{\prime}}_{\mu \nu\>(h)}\\
&+\frac{1}{2\pi\hat \alpha_{\text{GUT}}}\text{Re}(f_1) \lambda^{\dag a}_{(o)} {\slashed{\partial}}\lambda_{(o)}^{ a}  -\tfrac{i}{8\pi\hat \alpha_{\text{GUT}}}\text{Im}(f_{1})
{\partial}_\mu\left[ \lambda^{\dag a}_{(o)}\gamma_5\gamma^\mu \lambda^{ a}_{(o)}\right] \\
&+\frac{1}{2\pi\hat \alpha_{\text{GUT}}}\text{Re}(f_2) \lambda^{\dag a^{\prime}}_{(h)}{\slashed{\partial}}\lambda_{(h)}^{ a^{\prime}}     
-\tfrac{i}{8\pi\hat \alpha_{\text{GUT}}}\text{Im}(f_{2})
{\partial}_\mu\left[ \lambda^{\dag a^{\prime}}_{(h)}\gamma_5\gamma^\mu \lambda_{(h)}^{ a^{\prime}}\right]\ .
\end{split}
\end{equation}
Using $\text{Re}(f_1)=\text{Re}(f_2)=s$, expanding $s\rightarrow \langle s\rangle +\delta s$ around the fixed VEV $\langle s \rangle$, and using the fact that
\begin{equation}
\label{new1}
\hat \alpha_{\text{GUT}}=\alpha_{u} \langle s \rangle \ ,
\end{equation}
where $\alpha_{u}$ is the unification value of the gauge coupling parameter, we  obtain the following couplings of the vector bosons and gauginos to the real scalar component of the dilaton field:
\begin{equation}
\label{coat2}
\begin{split}
\mathcal{L}\supset& -\frac{1}{4\pi \alpha_{u}} F_{(o)}^{\mu \nu\> a}F^a_{\mu \nu\>(o)} -\frac{1}{4\pi \alpha_{u}}F_{(h)}^{\mu \nu\> a^{\prime}}F^{a^{\prime}}_{\mu \nu\>(h)}\\
& -\frac{1}{4\pi \alpha_{u}}\frac{1}{\langle s \rangle}\delta s F_{(o)}^{\mu \nu\> a}F^a_{\mu \nu\>(o)} -\frac{1}{4\pi \alpha_{u}}\frac{1}{\langle s \rangle}\delta s F_{(h)}^{\mu \nu\> a^{\prime}}F^{a^{\prime}}_{\mu \nu\>(h)} \\
\end{split}
\end{equation}
and
\begin{equation}
\label{coat3}
\begin{split}
\mathcal{L}\supset& +\frac{1}{2\pi \alpha_{u}} \lambda^{\dag a}_{(o)} {\slashed{\partial}}\lambda_{(o)}^{ a}+\frac{1}{2\pi  \alpha_{u}}\lambda^{\dag {a^{\prime}}}_{(h)}{\slashed{\partial}}\lambda_{(h)}^{ a^{\prime}}\\
& +\frac{1}{2\pi \alpha_{u}} \frac{1}{\langle s \rangle}\delta s\lambda^{\dag a}_{(o)}{\slashed{\partial}}\lambda_{(o)}^{ a}+\frac{1}{2\pi \alpha_{u}}\frac{1}{\langle s \rangle} \delta s \lambda^{\dag a}_{(h)}{\slashed{\partial}}\lambda_{(h)}^{ a^{\prime}} \\
\end{split}
\end{equation}
Furthermore, using $\text{Im}(f_1)=\text{Im}(f_2)=\sigma$, expanding $\sigma \rightarrow \langle \sigma \rangle +\delta \sigma$ around the fixed VEV $\langle \sigma\rangle$, and using \eqref{new1}, we find the following couplings of the vector bosons and gauginos to the axionic component of the dilaton field:
\begin{equation}
\label{coat4}
\begin{split}
\mathcal{L}\supset& -\frac{1}{4\pi \alpha_{u}} \frac{\langle \sigma \rangle}{\langle s \rangle} F_{(o)}^{\mu \nu\> a}\tilde{F}^a_{\mu \nu\>(o)} -\frac{1}{4\pi \alpha_{u}}  \frac{\langle \sigma \rangle}{\langle s \rangle} F_{(h)}^{\mu \nu\> a^{\prime}} \tilde{F}^{a^{\prime}}_{\mu \nu\>(h)}\\
& -\frac{1}{4\pi \alpha_{u}}\frac{1}{\langle s \rangle}\delta \sigma F_{(o)}^{\mu \nu\> a}\tilde{F}^a_{\mu \nu\>(o)} -\frac{1}{4\pi \alpha_{u}}\frac{1}{\langle s \rangle}\delta \sigma F_{(h)}^{\mu \nu\> a^{\prime}}\tilde{F}^{a^{\prime}}_{\mu \nu\>(h)}\\
\end{split}
\end{equation}
and
\begin{equation}
\label{coat5}
\begin{split}
\mathcal{L}\supset& +\frac{1}{8\pi \alpha_{u}}  \frac{\langle \sigma \rangle}{\langle s \rangle} {\partial}_\mu\left[ \lambda^{\dag a}_{(o)}  \gamma_{5} \gamma^{\mu} \lambda_{(o)}^{ a}\right]+\frac{1}{8\pi  \alpha_{u}} \frac{\langle \sigma \rangle}{\langle s \rangle}{\partial}_\mu\left[  \lambda^{\dag a^{\prime}}_{(h)}  \gamma_{5} \gamma^{\mu} \lambda_{(h)}^{ a^{\prime}} \right]\\
& +\frac{1}{8\pi \alpha_{u}} \frac{1}{\langle s \rangle}\delta \sigma {\partial}_\mu\left[ \lambda^{\dag a}_{(o)}  \gamma_{5} \gamma^{\mu} \lambda_{(o)}^{ a} \right]+\frac{1}{8\pi \alpha_{u}}\frac{1}{\langle s \rangle} \delta \sigma {\partial}_\mu\left[ \lambda^{\dag a^{\prime}}_{(h)}  \gamma_{5} \gamma^{\mu}  \lambda_{(h)}^{ a^{\prime}} \right]  \ .\\
\end{split}
\end{equation}

It is important to note that, in general, $\delta t$, $\delta s$, and $\delta \sigma$ are not the physical moduli mass eigenstates.
To proceed, one must re-express the above interactions in terms of these physical states. The moduli mass eigenstates, and their relation to  $\delta t$, $\delta s$ and $\delta \sigma$, depend strongly on the hidden sector gauge bundle and, in particular, on whether it does, or does not, contain an anomalous $U(1)$ factor. Each of these possibilities will be addressed in the next two subsections.

\subsection{Moduli Mass Eigenstates--Anomalous U(1) Case}

In the case that the hidden bundle structure group $G^{(2)}$ is an anomalous $U(1)$ Abelian group, then $H^{(2)}$ takes the form $H^{(2)}= \mathcal{H}_2 \times U(1)$. Hidden sectors with anomalous $U(1)$ factors have been studied in \cite{Anderson:2009nt,Anderson:2011cza,Weigand:2006yj,Ashmore:2020wwv}. 
Several of these admissible hidden sectors have gauge bundles which contain an ``anomalous'' $U(1)$ factor \cite{Dine:1987xk,Dine:1987gj,Anastasopoulos:2006cz}. 
For example, this has been accomplished within the context of the so-called $B-L$ MSSM theory ~\cite{Ambroso:2009jd,Marshall:2014kea,Marshall:2014cwa,Ovrut:2012wg,Ovrut:2014rba,Barger:2008wn,FileviezPerez:2009gr}, using only a single line bundle in the hidden sector \cite{Ashmore:2020ocb,Ashmore:2020wwv} It follows that the dilaton $S$ and K\"ahler modulus $T$ transform inhomogeneously as~\cite{Dumitru:2022apw,Dumitru:2021jlh}
\begin{equation}
\begin{split}
\label{eq:Killing_vects}
\delta_\theta S&=2i\pi a\epsilon_S^2\epsilon_R^2 \beta l \theta ~ \equiv k_S\theta\ ,\\
\delta_\theta T& =-2i a\epsilon_S\epsilon_R^2 l\theta  ~\equiv k_T\theta\ ,\\
\end{split}
\end{equation}
where the parameter $a$ depends on the line bundle embedding into the hidden $E_8$, while $\epsilon_{S}$ and $\epsilon_{R}$ are expansion parameters in the strong coupling regime.
Now assuming that $S$ and $T$ moduli have been stabilized at real VEVs $\langle s \rangle$ and $\langle t \rangle$ respectively and that these VEVs satisfy the condition that the Fayet-Iliopoulos (FI) term vanishes, as discussed in~\cite{Dumitru:2022apw,Dumitru:2021jlh}. Then the fluctuations $\delta S$ and $\delta T$ around these VEVs are related to the moduli mass eigenstates $\xi_{1}$ and $\xi_{2}$ by the linear relation
\begin{equation}
\label{light1}
\left(\begin{matrix}
\delta S\\\delta T
\end{matrix}
\right)={\mathbf U}^{-1}\left(\begin{matrix}
\xi^1\\\xi^2
\end{matrix}
\right)
\ ,
\end{equation}
where
\begin{align}
{\mathbf U}^{-1}&=
\frac{1}{\langle \Sigma \rangle}\begin{pmatrix}[1.4]
 k_S &\quad\sqrt{\langle \frac{g_{T\bar T}}{g_{S\bar S}}\rangle} k_T \\
 k_T &\quad-\sqrt{\langle \frac{g_{S\bar S}}{g_{T\bar T}}\rangle}  k_S 
\end{pmatrix} 
\label{eq:U22A}
\end{align}
and
\begin{equation}
\label{pad3}
\Sigma^2=g_{S\bar S}k_S\bar k_S+g_{T\bar T}k_T\bar k_T\  \ .
\end{equation}
The moduli metrics are $g_{S\bar S} =\partial_{S} \partial_{\bar S} K_{S}$ and $g_{T\bar T} =\partial_{T} \partial_{\bar T} K_{T}$,  where $K_{S}$ and $K_{T}$ are given in \eqref{burt6}.
As shown in \cite{Dumitru:2022apw,Dumitru:2021jlh}, the modulus eigenstate $\xi^{1}$ has an anomalous mass $m_{\text{anom}}\sim \mathcal{O}(M_U)$, where $M_{U}=3.15 \times 10^{16}$ GeV. Hence, it is very heavy and can, to leading order, be integrated out of the low-energy effective Lagrangian. This process is called $D$-term moduli stabilization in the presence of an anomalous $U(1)$. On the other hand, prior to $N=1$ supersymmetry breaking, the dimension one field $\xi^{2}$ is canonically normalized and massless. It then follows from \eqref{light1} and \eqref{eq:U22A} that 
\begin{equation}
\label{riv1}
\delta S=\frac{1}{\langle \Sigma \rangle}  {\langle  \frac{g_{T\bar T}}{g_{S\bar S}}\rangle}  k_T  \xi^{2}
\end{equation}
and 
\begin{equation}
\label{riv2}
\delta T=-\frac{1}{\langle \Sigma \rangle}  {\langle  \frac{g_{S\bar S}}{g_{T\bar T}}\rangle}  k_S  \xi^{2} \ .
\end{equation}
Writing 
\begin{equation}
\label{riv3}
\xi^{2}=\eta^{2}+i\phi^{2} \ ,
\end{equation}
one can then show using \eqref{burt6}, \eqref{eq:Killing_vects} and the constraint $FI=0$ that
\begin{equation}
\label{riv4}
\frac{1}{\langle t \rangle} \delta t = \frac{2}{\sqrt{3}} \frac{1} {\left(1+\frac{\beta^{2}}{3} \right)^{1/2}} \kappa_4\phi^{2},  \qquad  \frac{1}{\langle s \rangle} \delta s =\kappa_4 \phi^{2}, \qquad   \frac{1}{\langle s \rangle}\delta \sigma= - \kappa_4\eta^{2} \ ,
\end{equation}
where 
\begin{equation}
\label{rev5}
\beta=\frac{\langle s \rangle}{\langle t \rangle} \frac{3}{\pi \epsilon_{s}} \ .
\end{equation}

The real fields $\phi^{2}$ and $\eta^{2}$ are canonically normalized and, before spontaneously breaking supersymmetry, are massless. It is well known that within the context of heterotic string theories, spontaneously breaking $N=1$ supersymmetry via some combination of gaugino condensation, non-vanishing flux, and five-brane instantons can generate a non-vanishing hidden sector superpotential $\hat{W}$ and, hence, an F-term potential energy $V_{F}$ \cite{Dumitru:2022apw,Cicoli:2013rwa,Barreiro:1998nd,Binetruy:1996uv,Anderson:2011cza}. Since the supersymmetry breaking scale, $m_{SUSY}$ is defined as
\begin{equation}
\label{wall1}
m_{SUSY}=\kappa_{4}^{2}e^{\kappa_{4}^{2} \langle K_{S}+K_{T} \rangle /2} \langle |\hat{W}| \rangle \ ,
\end{equation}
it follows that, generically, after supersymmetry breaking both $\phi^{2}$ and $\eta^{2}$ remain mass eigenstates with
\begin{equation}
\label{riv6}
m_{\phi^{2}} \sim m_{\eta^{2}} \simeq {\cal{O}}(m_{SUSY}) \ .
\end{equation}
More explicitly, the exact expressions for $m_{\phi^{2}}$ and $m_{\eta^{2}}$, obtained by minimizing the complete potential energy $V=V_{D}+V_{F}$ assuming that $FI=0$, were given in \cite{Dumitru:2022apw}. These can be simplified using the fact that, to the lowest order in $\kappa_{4}^{2/3}$, the superpotential $\hat{W}$ generated by gaugino condensation and flux is generically a function of $S$ only. It then follows that
\begin{equation}
\label{pad1}
m_{\phi^{2}}^{2}= m_{ss}^{2} \langle \frac{g_{T\bar{T}}}{g_{S\bar{S}}} \frac{k_{T}k_{\bar{T}}}{\Sigma^{2}}\rangle \qquad,  \qquad m_{\eta^{2}}^{2}= m_{\sigma \sigma}^{2} \langle \frac{g_{T\bar{T}}}{g_{S\bar{S}}} \frac{k_{T}k_{\bar{T}}}{\Sigma^{2}} \rangle
\end{equation}
where
\begin{equation}
\label{pad2}
m_{ss}^{2}= \frac{\kappa_{4}^{2}}{2}\langle \frac{\partial^{2}V_{F}}{\partial s \partial s} \rangle \quad, \quad  m_{\sigma \sigma}^{2}= \frac{\kappa_{4}^{2}}{2}\langle \frac{\partial^{2}V_{F}}{\partial \sigma \partial \sigma} \rangle \ .
\end{equation}
Using \eqref{burt6} and \eqref{eq:Killing_vects}, the expressions in \eqref{pad1} simplify to
\begin{equation}
\label{wall3}
m_{\phi^{2}}^{2}= m_{ss}^{2}\langle s \rangle ^{2} \quad, \quad m_{\eta^{2}}^{2}= m_{\sigma \sigma}^{2}\langle s \rangle ^{2} \ .
\end{equation}
To proceed further, it is necessary to introduce an explicit non-perturbative superpotential and the associated $V_{F}$ potential. Let us choose the example presented in \cite{Dumitru:2022apw}, in which the K\"ahler potentials are identical to \eqref{burt6} and the non-perturbative superpotential on the hidden sector was taken to be
\begin{equation}
\label{cup1}
\hat{W}=M_{U}^{3}\left(c+e^{-bS}\right ) \ ,
\end{equation}
with c and b real numbers.
This leads to F-term potential energy of the form
\begin{equation}
\label{cup2}
V_{F}=\kappa^{2}_{4}M_{U}^{6}\frac{1}{16st^{3}}\left( |c+(2bs+1)e^{-b(s+i\sigma)}|^{2} \right) \ .
\end{equation}
Choosing, for example, the explicit parameters $b=3/4$ and  $c=1$, it was shown that the three relevant real moduli $s$, $t$ and $\sigma$ are stabilized at values
\begin{equation}
\label{wall2}
\langle s \rangle= \langle t \rangle \approx 2,  \quad \langle \sigma \rangle = \frac{4\pi}{3} \ .
\end{equation}
Evaluating $m_{ss}$ and $m_{\sigma \sigma}$ using \eqref{pad2}, \eqref{cup2} and \eqref{wall2}, and using expression \eqref{wall1} for $m_{SUSY}$, we find that
\begin{equation}
\label{wall4}
m_{\phi^{2}}=.431~ m_{SUSY} \quad, \quad m_{\eta^{2}}= 1.824~ m_{SUSY} \ ,
\end{equation}
which are consistent with the generic expression \eqref{riv6} .

For an anomalous $U(1)$ hidden sector, the Lagrangian couplings  shown in \eqref{eq:int_sc_1} and \eqref{eq:int_sc_2} generate the following types of vertex amplitudes.

\begin{itemize}

\item{matter scalars - $\phi^2$ modulus}:

\begin{align}
\label{scalar_t}
\raisebox{-36pt}{\includegraphics[]{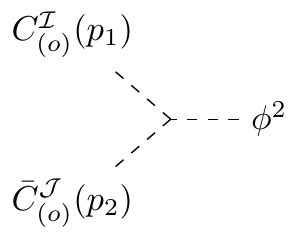}}
&=\frac{2}{\sqrt{3}} \frac{1} {(1+\frac{\beta^{2}}{3} )^{1/2}}\delta_{{\cal I}\bar {\cal J}}\>p_{1\mu}p_2^{\mu}\> \kappa_4\ ,\\
\raisebox{-36pt}{\includegraphics[]{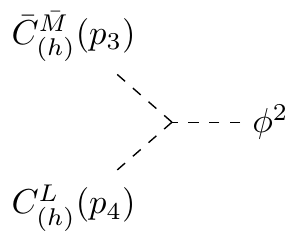}}
&=\frac{2}{\sqrt{3}} \frac{1} {(1+\frac{\beta^{2}}{3} )^{1/2}}\delta_{{L}\bar {M}}\>p_{3\mu}p_4^{\mu}\>{\kappa_4}\ .
\end{align}

\item{matter fermions - $\phi^2$ modulus: }

\begin{align}
 \raisebox{-36pt}{\includegraphics[]{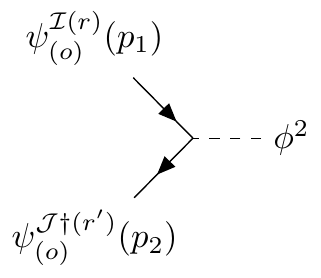}}
&=\frac{2}{\sqrt{3}} \frac{1} {(1+\frac{\beta^{2}}{3} )^{1/2}}
\delta_{{\cal I}\bar {\cal J}}\>{\kappa_4M_{\Psi^{\cal I}_{(o)}}}\bar u^{r}(p_1)v^{r^\prime}(p_2)\ ,\\
\raisebox{-36pt}{\includegraphics[]{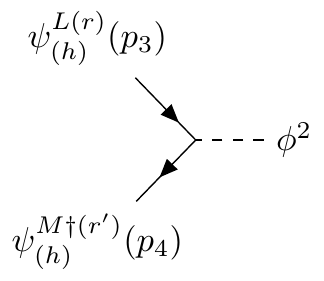}}
&
=\frac{2}{\sqrt{3}} \frac{1} {(1+\frac{\beta^{2}}{3} )^{1/2}}\delta_{L\bar M}\>{\kappa_4M_{\Psi^{L}_{(h)}}}\bar u^{r}(p_3)v^{r^\prime}(p_4)\ .
\end{align}

\end{itemize}
In the above, $p_1$ and $p_2$ are the momenta of the observable sector particles, while $p_3$
and $p_4$ respresent the momenta of the hidden sector particles. Furthermore, $r$ and $ r^\prime$ indices 
represent the two spin states of each fermion.
Note that we have used  $\bar u^{r}(p_1)(\slashed{p_1}-\slashed{p_2})v^{r^\prime}(p_2)=M_{\Psi^{\cal I}_{(o)}}{\langle t\rangle }\bar u^{r}(p_1)v^{r^\prime}(p_2)$ to obtain the vertex amplitudes from above.
Here $M_{\Psi^{{\cal {I}}(r)}_{(o)}}$ is the time-dependent fermion mass defined in \eqref{A15} of Appendix A, while $u^r(p)$ and $v^r(p)$ are Dirac four-spinors with the properties
\begin{equation}
\sum_{{r}=1}^2u^r\bar u^r= \slashed{p}+m\ ,\quad \sum_{{r}=1}^2v^r\bar v^r= \slashed{p}-m\ ,
\end{equation}
where the summation is over all the possible spin states.

For the case in which the hidden sector structure group contains an anomalous $U(1)$ factor, the Lagrangian couplings  shown in \eqref{coat2} and \eqref{coat3} generate the following types of vertex amplitudes. Using the fact, discussed above, that $ \delta s =\kappa_4 {\langle s \rangle} \phi^{2}$, it follows that

\begin{itemize}

\item{gauge fields - $\phi^2$ modulus:}

\begin{align}
 \raisebox{-36pt}{\includegraphics[]{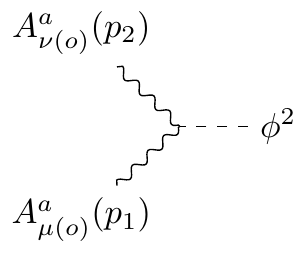}}
&=-\eta_{\mu \nu}p_1^\mu p_2^\nu\>{\langle s \rangle}\frac{\kappa_4}{\pi\hat \alpha_{\text{GUT}}}\ ,\\
 \raisebox{-36pt}{\includegraphics[]{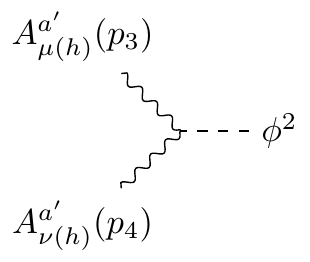}}
&=-\eta_{\mu \nu}p_3^\mu p_4^\nu\>{\langle s \rangle}\frac{\kappa_4}{\pi\hat \alpha_{\text{GUT}}}\ .
\end{align}

\item{gauginos - $\phi^2$ modulus:}

\begin{align}
 \raisebox{-36pt}{\includegraphics[]{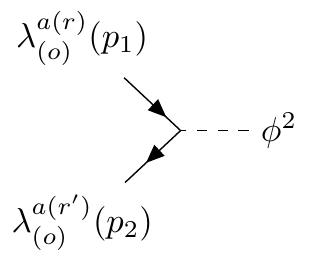}}
&\nonumber=\frac{\kappa_4}{2\pi\hat \alpha_{\text{GUT}} }\bar u^r(p_1)(\slashed{p_1}-\slashed{p_2})v^{r^\prime}(p_2)\\
&=
{\langle s \rangle}\frac{\kappa_4M_{\lambda_{(o)}^a}}{2\pi\hat \alpha_{\text{GUT}}}\bar u^r(p_1)v^{r^\prime}(p_2)\ 
\label{gaugino_s1} \ , \\
 \raisebox{-36pt}{\includegraphics[]{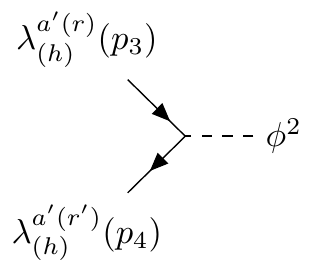}}
&\nonumber=\frac{\kappa_4}{2\pi\hat \alpha_{\text{GUT}} }\bar u^r(p_3)(\slashed{p_3}-\slashed{p_4})v^{r^\prime}(p_4)\\
&=
{\langle s \rangle}\frac{\kappa_4M_{\lambda_{(h)}^{a^\prime}}}{2\pi\hat \alpha_{\text{GUT}}}\bar u^r(p_3)v^{r^\prime}(p_4)\ .
\label{gaugino_s2}
\end{align}

\end{itemize}
Here $M_{{\lambda}_{(o)}^a}$ and $M_{{\lambda}_{(h)}^{a^{\prime}}}$ are the gaugino masses in the observable and hidden sectors respectively. In the hidden sector, the gaugino masses originate in the soft-SUSY breaking Lagrangian and are fixed constants. However, as discussed in Appendix A, 
in the observable sector they are a combination of soft-SUSY breaking masses and--during the reheating phase-- the masses generated by the time-dependent inflaton VEV$\sqrt{\langle\psi^2 \rangle }$.

Unlike the matter multiplets, the gauge multiplets also couple to the axionic component of the dilaton. For the case in which the hidden sector structure group contains an anomalous $U(1)$ factor, the Lagrangian couplings  shown in \eqref{coat4} and \eqref{coat5} generate the following types of vertex amplitudes. Using the fact, discussed above, that $\delta \sigma = -\kappa_4 \langle s \rangle \eta^{2}$, it follows that the observable and hidden sector fields couple to the axion component $\eta^2$ in the following way
\begin{itemize}

\item{observable gauge fields/gauginos - $\eta^2$ axion:}

\begin{align}
\centering
  \raisebox{-36pt}{\includegraphics[]{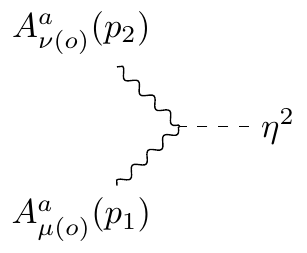}}
&=-\eta_{\mu \nu}p_1^\mu p_2^\nu\>{\langle s \rangle}\frac{\kappa_4}{\pi\hat \alpha_{\text{GUT}}}\ ,\\
 \raisebox{-36pt}{\includegraphics[]{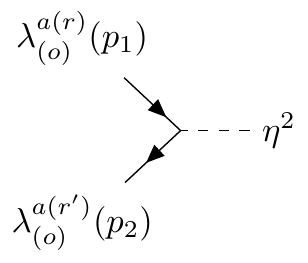}}
&\nonumber=\frac{\kappa_4}{2\pi\hat \alpha_{\text{GUT}} }\bar u^r(p_1)(\slashed{p_1}-\slashed{p_2})v^{r^\prime}(p_2)\\
&=
{\langle s \rangle}\frac{\kappa_4M_{\lambda_{(o)}^a}}{2\pi\hat \alpha_{\text{GUT}}}\bar u^r(p_1)v^{r^\prime}(p_2)\ 
\label{gaugino_s1} \ .
\end{align}

\item{hidden gauge fields/gauginos -  $\eta^2$ axion:}

\begin{align}
\centering
  \raisebox{-36pt}{\includegraphics[]{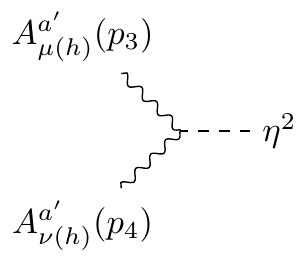}}
&=-\eta_{\mu \nu}p_3^\mu p_4^\nu\>{\langle s \rangle}\frac{\kappa_4}{\pi\hat \alpha_{\text{GUT}}}\ ,\\
 \raisebox{-36pt}{\includegraphics[]{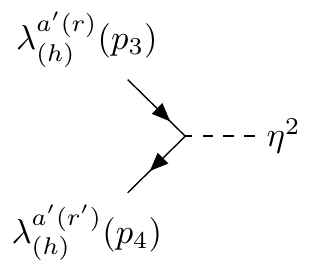}}
&\nonumber=\frac{\kappa_4}{2\pi\hat \alpha_{\text{GUT}} }\bar u^r(p_3)(\slashed{p_3}-\slashed{p_4})v^{r^\prime}(p_4)\\
&=
{\langle s \rangle}\frac{\kappa_4M_{\lambda_{(h)}^{a^\prime}}}{2\pi\hat \alpha_{\text{GUT}}}\bar u^r(p_3)v^{r^\prime}(p_4)\ 
\label{gaugino_s1extra} \ .
\end{align}
\end{itemize}


\subsection{Moduli Mass Eigenstates--Non-Anomalous Case}

When the hidden sector bundle structure group $G^{(2)}$ does {\it not} contain an anomalous U(1) factor, the low energy gauge group can be reduced to the form $H^{(2)}= \mathcal{H}_2$, where either all of $\mathcal{H}_2$, or one or more non-Abelian subfactors, can contribute to gaugino condensation. This situation is fundamentally different than the anomalous $U(1)$  case discussed in the previous subsection, since there is no longer an anomalous mass $m_{\text{anom}}\sim \mathcal{O}(M_U)$ generated which eliminates two real scalar moduli degrees of freedom. That is, in the non-anomalous case one must consider all four real degrees of freedom in the dilaton $S$ and K\"ahler modulus $T$.

 As discussed in Appendix B, spontaneously breaking $N=1$ supersymmetry via some combination of gaugino condensation, non-vanishing flux and five-brane instantons can generate a non-vanishing hidden sector superpotential $\hat{W}$ and, hence, an F-term potential energy $V_{F}$. This can be chosen so as to fix the VEVs for $\langle s \rangle$, $\langle \sigma \rangle$, $\langle t \rangle$ and $\langle \chi \rangle$. We begin by expanding 
 \begin{equation}
 \label{bang1}
s= \langle s \rangle+\delta s\quad, \quad t= \langle t \rangle+\delta t \ .
 \end{equation}
As discussed in Appendix B, when $s$ and $t$ are expanded around their VEVs, neither $\delta s$ nor $\delta t$ are the mass eigenstates. These are given by the linear combinations
\begin{equation}
\begin{split}
\label{eq:eigen_st}
\delta \tilde s&=x_1\kappa_4^{-1}\delta s+x_2\kappa_4^{-1}\delta t\ ,\\
\delta \tilde t&=y_1\kappa_4^{-1}\delta s+y_2\kappa_4^{-1}\delta t\ ,
\end{split}
\end{equation}
where the parameters $x_1,x_2,y_1,y_2$ are dimensionless. Therefore,
in order to find the vertex interactions between the matter fields and the physical moduli mass eigenstates, given in \eqref{eq:int_sc_1} and \eqref{eq:int_sc_2}, and the vertex interactions between the gauge fields and the physical moduli mass eigenstates, given in \eqref{coat2} and \eqref{coat3}, one needs to rotate $\delta s$ and $\delta t$ into $\delta \tilde s$ and $\delta \tilde t$, using the inverse of  \eqref{eq:eigen_st}.
This rotation introduces some
arbitrariness into our analysis since one cannot determine the rotation parameters  $x_1,x_2,y_1,y_2$ in the absence of a specific 
non-perturbative superpotential that stabilizes the moduli. However, as long as they are of order one--which the example below indicates that they are--their specific values cannot have a significant impact on the estimation of the dark matter relic density expected in our theory. A detailed analysis of what values these rotation parameters can have is therefore unnecessary for this work.

Be that as it may, we now present a toy model for moduli stabilization which indicates not only that the parameters $x_1,x_2,y_1,y_2$ are not large, but that one can expect $\delta \tilde{s} \simeq \kappa_4^{-1}\delta s$ and 
$\delta \tilde{t} \simeq \kappa_4^{-1}\delta t$.  Using the formalism presented in Appendix B, it follows that in models with no contributions from five-brane non-perturbative effects, the non-perturbative superpotential
generated by gaugino condensates and flux is given, to order $\kappa_4^{2/3}$, by 
  \begin{equation}
  \label{coat1}
  \hat{W}(S,T) = M_{U}^{3} \big(c+\sum_{i} e^{-b_{i}S} \big) \ \ ,
  \end{equation}
  where the index $i$ sums over factor subgroups ${\cal{H}}^{r}_{2}\subset {\cal{H}}_{2}$ which contribute to gaugino condensation.
Note that this has no dependence on the $T$ modulus. It follows that the F-term potential generated by such a superpotential can stabilize the dilaton $S$ components and lead to mass terms for the $\kappa_4^{-1}\delta s$ and $\kappa_4^{-1}\delta \sigma$ states, while leaving the $t$ and $\chi$ directions of $T$ flat. In this case, 
  the moduli mass matrix is diagonal--with non-zero masses for  $\delta s$ and $\delta \sigma $ and zero mass eigenvalues for $\delta t$ and $\delta \chi$ respectively. That is, 
  the moduli eigenstates are exactly $\delta \tilde s=\kappa_4^{-1} \delta s$, $\delta \tilde \sigma=\kappa_4^{-1} \delta \sigma$ and $\delta \tilde t=\kappa_4^{-1} \delta t$,  $\delta \tilde \chi=\kappa_4^{-1} \delta \chi$. 
However, the cosmological moduli problem \cite{Kusenko:2012ch} tells us that massless moduli eigenstates cannot be allowed to exist in the theory. Otherwise too many would be produced during inflation and would overdilute the universe. This problem can be solved by turning on the small genus-one corrections in the theory. Such order $\kappa_4^{4/3}$terms introduce a weak $T$-dependence into the superpotential of the type~\cite{Lukas:1997fg}
  \begin{equation}
\label{eq:Wst23}
  {\hat{W}}(S,T)=M_{U}^{3} \big(c+\sum_i e^{-b_{i}(S+\beta_{1} T)} \big)\ 
  \end{equation}
 with coefficient $\beta_{1} \sim {\cal{O}}(\kappa_4^{4/3}) \ll 1$. This correction lifts the flatness in the $t$ and $\chi$ directions. Assuming stabilization can be demonstrated within this context, these genus-one corrections now mix $\delta s$ and $\delta t$ into new mass eigenstates.
 Specifically, to linear order in $\beta_{1}$, the $\delta \tilde s$ and $\delta \tilde t$ moduli mass eigenstates are now given by expressions \eqref{eq:eigen_st} where
%
\begin{equation}
x_1,\>y_2\sim \mathcal{O}(1) \quad  {\rm and} \quad  x_2,\>y_1\sim \mathcal{O}(\beta_{1}) \ .
\label{paper1}
\end{equation}
Hence
\begin{equation}
\delta \tilde s \simeq \kappa_4^{-1}\delta s  \quad  {\rm and} \quad \delta \tilde t \simeq \kappa_4^{-1}\delta t \ ,
\label{paper2}
\end{equation}
as expected.
Furthermore, as discussed in Appendix B, the associated mass eigenvalues are  
\begin{equation}
m_{\tilde s}\sim m_{\text{SUSY}}\ , \quad m_{\tilde t}\sim \beta_{1} m_{\text{SUSY}}\ ,
\end{equation}
thus solving part of the cosmological moduli problem. 

Let us now expand
 \begin{equation}
 \label{bang2}
\sigma= \langle \sigma \rangle+\delta \sigma\quad, \quad \chi= \langle \chi \rangle+\delta \chi \ .
 \end{equation}
As with $s$ and $t$ above, when $\sigma$ and $\chi$ are expanded around their VEVs neither $\delta \sigma$ nor $\delta \chi$ are the mass eigenstates. These are given by the linear combinations
\begin{equation}
\begin{split}
\label{eq:eigen_st2}
\delta  \tilde \sigma&=x_1^\prime\kappa_4^{-1}\delta \sigma+x_2^\prime\kappa_4^{-1}\delta\chi\ ,\\
 \delta \tilde \chi&=y_1^\prime\kappa_4^{-1}\delta \sigma+y_2^\prime\kappa_4^{-1}\delta\chi\ ,
\end{split}
\end{equation}
where the parameters $x_1^\prime,x_2^\prime,y_1^\prime,y_2^\prime$ are dimensionless.
In order to find the vertex interactions between the gauge fields and the physical moduli mass eigenstates, given in \eqref{coat4} and \eqref{coat5}, one needs to rotate $\delta \sigma$ and $\delta \chi$ into $\delta \tilde \sigma$ and $\delta \tilde \chi$ using the inverse of  \eqref{eq:eigen_st2}.
Again employing the superpotential in \eqref{eq:Wst23} to linear order in $\beta_{1}$, the $\delta \tilde \sigma$ and $\delta \tilde \chi$ moduli mass eigenstates are now given by expressions \eqref{eq:eigen_st2} where
\begin{equation}
x_1^{\prime},\>y_2^{\prime}\sim \mathcal{O}(1) \quad  {\rm and} \quad  x_2^{\prime},\>y_1^{\prime}\sim \mathcal{O}(\beta_{1}) \ .
\label{paper1A}
\end{equation}
Hence
\begin{equation}
\delta \tilde \sigma \simeq \kappa_4^{-1}\delta \sigma  \quad  {\rm and} \quad \delta \tilde \chi \simeq \kappa_4^{-1}\delta \chi\ .
\label{paper2A}
\end{equation}
Furthermore,
the associated mass eigenvalues are again
\begin{equation}
m_{\tilde \sigma}\sim m_{\text{SUSY}}\ , \quad m_{\tilde \chi}\sim \beta_{1} m_{\text{SUSY}}\ ,
\end{equation}
thus solving the cosmological moduli problem. 

Therefore, for the case in which the hidden sector structure group does not contain an anomalous $U(1)$ factor, the Lagrangian couplings  shown in \eqref{eq:int_sc_1} and \eqref{eq:int_sc_2} generate the following types of vertex amplitudes. Using the fact discussed above that one expects $\delta \tilde{t} \simeq \kappa_4^{-1}\delta t$, it follows that
\begin{itemize}

\item{matter scalars - $\delta \tilde t$ modulus}:

\begin{align}
\label{scalar_t}
\raisebox{-32pt}{\includegraphics[]{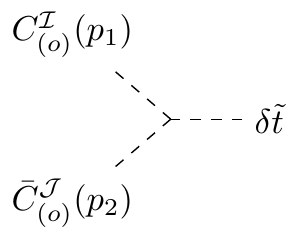}}=\delta_{{\cal I}\bar {\cal J}}\>p_{1\mu}p_2^{\mu}\>\frac{\kappa_4}{\langle t\rangle}\ ,
\raisebox{-32pt}{\includegraphics[]{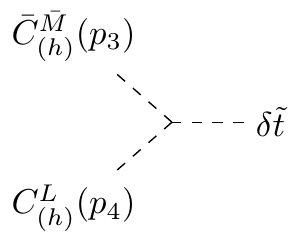}}=\delta_{{L}\bar {M}}\>p_{3\mu}p_4^{\mu}\>\frac{\kappa_4}{\langle t\rangle}\ .
\end{align}

\item{matter fermions  - $\delta \tilde t$ modulusi: }

\begin{align}
\raisebox{-37pt}{\includegraphics[]{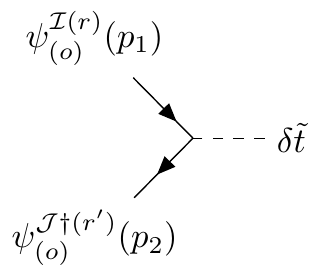}}
&=
\delta_{{\cal I}\bar {\cal J}}\>\frac{\kappa_4M_{\Psi^{\cal I}_{(o)}}}{\langle t\rangle }\bar u^{r}(p_1)v^{r^\prime}(p_2)\ ,\\
\raisebox{-37pt}{\includegraphics[]{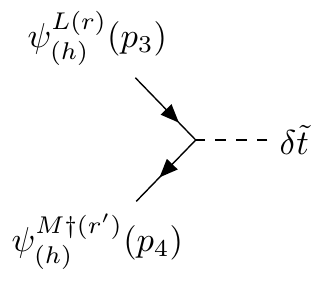}}
&=\delta_{L\bar M}\>\frac{\kappa_4M_{\Psi^{L}_{(h)}}}{\langle t\rangle }\bar u^{r}(p_3)v^{r^\prime}(p_4)\ .
\end{align}
\end{itemize}
Here $M_{\Psi^{{\cal {I}}(r)}_{(o)}}$ is the time-dependent fermion mass defined in \eqref{A15}, while $u^r(p)$ and $v^r(p)$ are Dirac four-spinors with the properties
\begin{equation}
\sum_{{r}=1}^2u^r\bar u^r= \slashed{p}+m\ ,\quad \sum_{{r}=1}^2v^r\bar v^r= \slashed{p}-m\ ,
\end{equation}
where the summation is over all the possible spin states.

For the case in which the hidden sector structure group does not contain an anomalous $U(1)$ factor, the Lagrangian couplings  shown in \eqref{coat2} and \eqref{coat3} generate the following types of vertex amplitudes. Using the fact discussed above that one expects $\delta \tilde{s} \simeq \kappa_4^{-1}\delta s$, it follows that

\begin{itemize}

\item{gauge boson fields  - $\delta \tilde s$ modulus:}

\begin{align}
\raisebox{-35pt}{\includegraphics[]{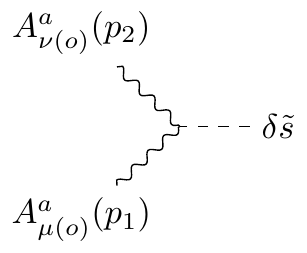}}
&=-\eta_{\mu \nu}p_1^\mu p_2^\nu\>\frac{\kappa_4}{\pi\hat \alpha_{\text{GUT}}}\ ,\\
\raisebox{-35pt}{\includegraphics[]{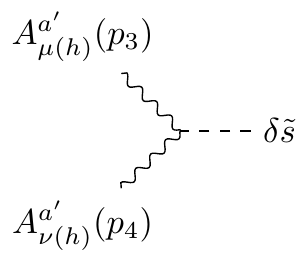}}
&=-\eta_{\mu \nu}p_3^\mu p_4^\nu\>\frac{\kappa_4}{\pi\hat \alpha_{\text{GUT}}}\ .
\end{align}

\item{gauginos  - $\delta \tilde s$ modulus:}

\begin{align}
\raisebox{-35pt}{\includegraphics[]{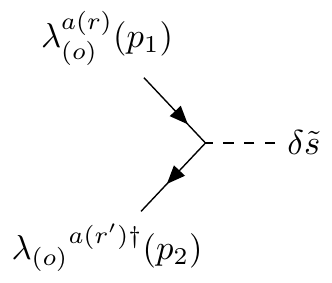}}
&=\frac{\kappa_4}{2\pi\hat \alpha_{\text{GUT}} }\bar u^r(p_1)(\slashed{p_1}-\slashed{p_2})v^{r^\prime}(p_2)=
\frac{\kappa_4M_{\lambda_{(o)}^a}}{2\pi\hat \alpha_{\text{GUT}}}\bar u^r(p_1)v^{r^\prime}(p_2)\ 
\label{gaugino_s1} \ , \\
\raisebox{-35pt}{\includegraphics[]{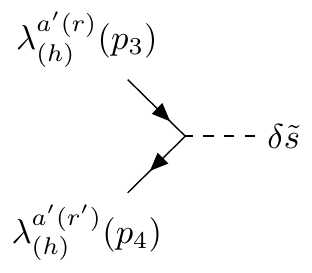}}
&=\frac{\kappa_4}{2\pi\hat \alpha_{\text{GUT}} }\bar u^r(p_3)(\slashed{p_3}-\slashed{p_4})v^{r^\prime}(p_4)=
\frac{\kappa_4M_{\lambda_{(h)}^{a^\prime}}}{2\pi\hat \alpha_{\text{GUT}}}\bar u^r(p_3)v^{r^\prime}(p_4)\ .
\label{gaugino_s2}
\end{align}
\end{itemize}
Here $M_{{\lambda}_{(o)}^a}$ and $M_{{\lambda}_{(h)}}^{a^\prime}$ are the gaugino masses in the observable and hidden sectors respectively. In the hidden sector, the gaugino masses originate in the soft-SUSY breaking Lagrangian and are fixed constants. However, as discussed in Appendix A, 
in the observable sector they are a combination of soft-SUSY breaking masses and--during the reheating phase-- the masses generated by the time-dependent inflaton VEV$\sqrt{\langle\psi^2 \rangle }$.

Unlike the matter multiplets, the gauge multiplets also couple to the axionic component of the dilaton. For the case in which the hidden sector structure group does not contain an anomalous $U(1)$ factor, the Lagrangian couplings  shown in \eqref{coat4} and \eqref{coat5} generate the following types of vertex amplitudes. Using the fact discussed above that one expects $\delta \tilde{\sigma} \simeq \kappa_4^{-1}\delta \sigma$, it follows that the observable and hidden sector fields couple to the axion component $\delta\tilde \sigma$ in the following way
\begin{itemize}
\item{observable gauge fields/gauginos - $\delta \tilde \sigma$ axion:}
\begin{align}
\centering
 \raisebox{-35pt}{\includegraphics[]{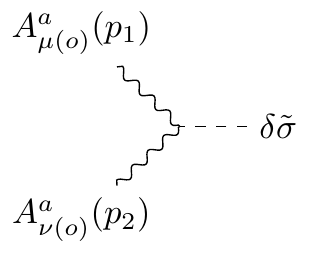}}
&=-\eta_{\mu \nu}p_1^\mu p_2^\nu\>\frac{\kappa_4}{\pi\hat \alpha_{\text{GUT}}}\ ,\\
\raisebox{-35pt}{\includegraphics[]{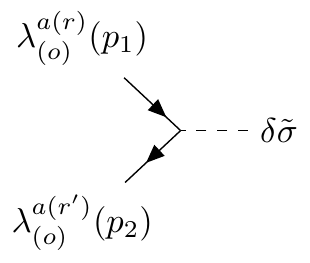}}
&=\frac{\kappa_4}{2\pi\hat \alpha_{\text{GUT}} }\bar u^r(p_1)(\slashed{p_1}-\slashed{p_2})v^{r^\prime}(p_2)=
\frac{\kappa_4M_{\lambda_{(o)}^a}}{2\pi\hat \alpha_{\text{GUT}}}\bar u^r(p_1)v^{r^\prime}(p_2)\ 
\label{gaugino_s1} \ .
\end{align}

\item{hidden gauge fields/gauginos - $\delta \tilde \sigma$ axion:}
\begin{align}
\centering
 \raisebox{-35pt}{\includegraphics[]{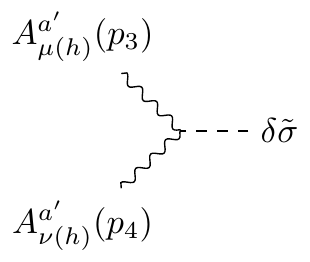}}
&=-\eta_{\mu \nu}p_1^\mu p_2^\nu\>\frac{\kappa_4}{\pi\hat \alpha_{\text{GUT}}}\ ,\\
\raisebox{-35pt}{\includegraphics[]{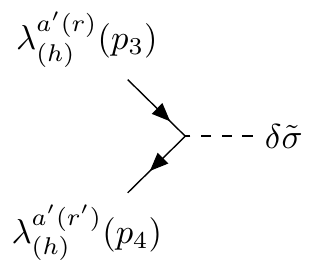}}
&=\frac{\kappa_4}{2\pi\hat \alpha_{\text{GUT}} }\bar u^r(p_3)(\slashed{p_3}-\slashed{p_3})v^{r^\prime}(p_4)=
\frac{\kappa_4M_{\lambda_{(h)}^{a^\prime}}}{2\pi\hat \alpha_{\text{GUT}}}\bar u^r(p_3)v^{r^\prime}(p_4)\ 
\label{gaugino_s1} \ .
\end{align}

\end{itemize}

\section{ Dark Matter}

In Section \ref{sec:modPortal}, we presented the dominant interactions that couple the fields in both the observable sector and the hidden sector to moduli scalars. These interactions are responsible for the ``moduli portals'' between two otherwise independent sectors. 
In this section, we will identify which hidden sector particles are the most likely candidates for dark matter. We will also propose a possible dark matter production mechanism and then verify that this mechanism can indeed produce the expected dark matter abundance we observe today. 
As discussed in detail in Section \ref{sec:modPortal}, these moduli portal interactions break into two distinct types: 1) those with an anomalous $U(1)$ factor in the hidden sector gauge group and 2) vacua without an anomalous $U(1)$ factor. Therefore, we must present our discussion of chiral dark matter production for each of these two types separately. 

\subsection{Dark Matter Production Mechanism--Anomalous U(1) Case}

The hidden sector fields are 
a promising dark matter candidate, since--other than the moduli portal interactions discussed above--they have no interactions with ``normal'' matter in the observable sector other than supergravitational. As discussed in Appendix C,
we find that the hidden matter fermions, as well as any non-anomalous hidden sector gauge bosons, remain  {\it massless} after $N=1$ supersymmetry is broken in the 4D vacuum. Since, as we will see below, all amplitudes for decays to dark matter are proportional to the dark matter mass, it follows that hidden matter fermions and hidden gauge bosons cannot be dark matter. Of course, 
there may exist an analog of the Higgs mechanism in the hidden sector 
which gives masses to the fermions and gauge bosons at low energies. However, 
we will assume this is not the case. It follows that {\it the only possible dark matter candidates are the hidden matter scalars and the hidden sector gauginos}. 

\subsubsection{${\phi}^{2}$ modulus portal:}

Using the results of subsection 3.3, the processes by which observable sector scalars, fermions, gauge fields, and gauginos can decay into hidden sector matter scalars and gauginos via the $\phi^{2}$ modulus are shown in Figure 1.
\begin{figure}[t]
   \centering
       \begin{subfigure}[b]{0.65\textwidth}
       \centering
   \includegraphics[width=1\textwidth]{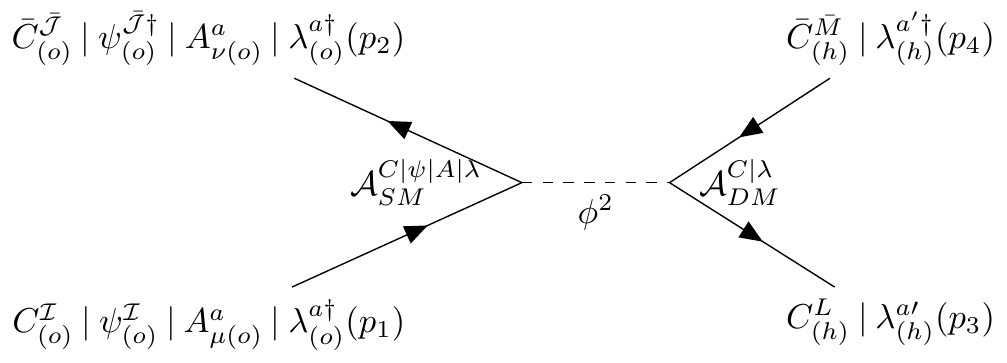}
\end{subfigure}\hfill
\caption{ Possible chiral dark matter production mechanisms via the $\phi^2$ modulus portal. The massive $\phi^2$ modulus can be produced by scalar-scalar, fermion-fermion, boson-boson and gaugino-gaugino scattering in the thermal bath of particles in the observable sector created during reheating. Based on the expected mass spectrum and the possible interactions between the observable sector thermal bath and the hidden sector fields, we identified the hidden matter scalars $C_{(h)}^L$, $L=1,\dots, {N}_{(h)}$, and hidden gauginos $\lambda^{a^{\prime}}_{(h)}$, as the possible dark matter candidates within our theoretical context. For the observable sector gauginos, the index $a$ runs over the adjoint representation of $SU(3)_{C} \times SU(2)_{L} \times U(1)_{Y}$ and any further extension of the standard model gauge group--such as $U(1)_{B-L}$ in the $B-L$ MSSM. For the hidden sector gauginos, the index $a^{\prime}$ runs over the adjoint representation of the hidden sector group $\mathcal{H}_1$ which does not
contribute to gaugino condensation.}
\label{fig:int_9}
\end{figure}
The strength of interaction vertices in Figure 1 for the couplings of the observable sector fields to the $\phi^2$ modulus were found in subsection 3.3 to be
\begin{align}
\mathcal{A}_{SM}^C&=\frac{2}{\sqrt{3}} \frac{1} {(1+\frac{\beta^{2}}{3} )^{1/2}}\delta_{{\cal I}\bar {\cal J}}\>p_{1\mu}p_2^{\mu}\> \kappa_4\equiv 
\delta_{{\cal I}\bar {\cal J}}\>p_{1\mu}p_2^{\mu}\>\kappa_4\alpha^C_{\text{SM}}\ ,\\
\mathcal{A}_{SM}^\psi&=\frac{2}{\sqrt{3}} \frac{1} {(1+\frac{\beta^{2}}{3} )^{1/2}}
\delta_{{\cal I}\bar {\cal J}}\>{\kappa_4M_{\Psi^{\cal I}_{(o)}}}\bar u^{r}(p_1)v^{r^\prime}(p_2)
=\delta_{{\cal I}\bar {\cal J}}\>{\kappa_4M_{\Psi^{\cal I}_{(o)}}}\bar u(p_1)v(p_2) \alpha_{\text{SM}}^\psi \ ,\\
\mathcal{A}_{SM}^A&=p_{1\mu}p_2^{\mu}\>{\langle s \rangle}\frac{\kappa_4}{\pi\hat \alpha_{\text{GUT}}}\equiv 
p_{1\mu}p_2^{\mu}\>\kappa_4\alpha^A_{\text{SM}}\ ,\\ 
\mathcal{A}_{SM}^\lambda&={\langle s \rangle}\frac{\kappa_4M_{{\lambda}_{(o)}}}{2\pi\hat \alpha_{\text{GUT}} }\bar u(p_1)v(p_2)
={\kappa_4M_{{\lambda}_{(o)}}}\bar u(p_1)v(p_2) \alpha_{\text{SM}}^\lambda \ .
\end{align}
Similarly, the strength of the interaction vertices in Figure 1 for the couplings of the hidden sector scalar and gaugino fields to the $\phi^2$ modulus were found to be
\begin{align}
\mathcal{A}_{DM}^C&=\frac{2}{\sqrt{3}} \frac{1} {(1+\frac{\beta^{2}}{3} )^{1/2}} \delta_{L\bar{M}} \>p_{3\mu}p_4^{\mu}\> \kappa_4\equiv 
\delta_{L\bar{M}} \>p_{3\mu}p_4^{\mu}\>\kappa_4\alpha^C_{\text{DM}}\ ,\\
\mathcal{A}_{DM}^\lambda&=\langle s \rangle\frac{\kappa_4M_{{\lambda}_{(h)}}}{2\pi\hat \alpha_{\text{GUT}} }\bar u(p_3)v(p_4)
={\kappa_4M_{{\lambda}_{(h)}}}\bar u(p_3)v(p_4) \alpha_{\text{DM}}^\lambda \ .
\end{align}

In the above, we defined the parameters
\begin{equation}
\label{eq:coupling_def11}
\alpha^C_{\text{SM}}=\alpha_{\text{SM}}^\Psi=\alpha^C_{\text{DM}}=\frac{2}{\sqrt{3}} \frac{1} {(1+\frac{\beta^{2}}{3} )^{1/2}}\ ,
\end{equation}
and
\begin{equation}
\label{eq:coupling_def22}
\tfrac{1}{2}\alpha^A_{\text{SM}}=\alpha_{\text{SM}}^\lambda=\alpha^\lambda_{\text{DM}}=\frac{\langle s\rangle }{2\pi\hat \alpha_{\text{GUT}}}\ .
\end{equation}

\subsubsection{${\eta}^{2}$ modulus portal:}

Using the results of subsection 3.3, the processes by which observable sector gauge fields and gauginos can decay into hidden sector gauginos via the $\eta^{2}$ modulus are shown in Figure 2.
\begin{figure}[t]
   \centering
       \begin{subfigure}[b]{0.53\textwidth}
       \centering
   \includegraphics[width=1\textwidth]{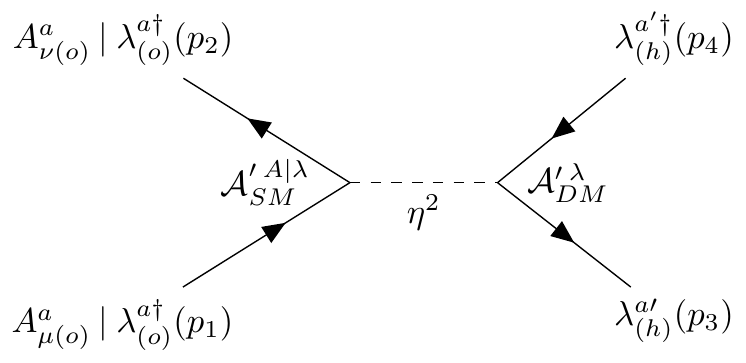}
\end{subfigure}\hfill
\caption{ Possible chiral dark matter production mechanisms via the $\eta^2$ modulus portal. The massive $\eta^2$ modulus can be produced either by boson-boson or gaugino-gaugino scattering in the thermal bath of particles in the observable sector created during reheating. The $\eta^{2}$ modulus can then decay to the hidden sector gauginos.}
\label{fig:int_10}
\end{figure}
The strength of interaction vertices in Figure 2 for the couplings of the observable sector fields to the $\eta^2$ modulus were found in subsection 3.3 to be
\begin{align}
{\mathcal{A}^{\prime}}_{SM}^A&=-p_{1\mu}p_2^{\mu}\>{\langle s \rangle}\frac{\kappa_4}{\pi\hat \alpha_{\text{GUT}}}\equiv 
p_{1\mu}p_2^{\mu}\>\kappa_4{\alpha^{\prime}}^A_{\text{SM}}\ ,\\ 
{\mathcal{A}^{\prime}}_{SM}^\lambda&={\langle s \rangle}\frac{\kappa_4M_{{\lambda}_{(o)}}}{2\pi\hat \alpha_{\text{GUT}} }\bar u(p_1)v(p_2)
={\kappa_4M_{{\lambda}_{(o)}}}\bar u(p_1)v(p_2) {\alpha^{\prime}}_{\text{SM}}^\lambda\ .
\end{align}
Similarly, the strength of the interaction vertices in Figure 2 for the couplings of the hidden sector gaugino fields to the $\eta^2$ modulus were found to be
\begin{align}
{\mathcal{A}^{\prime}}_{DM}^\lambda&=\langle s \rangle\frac{\kappa_4M_{{\lambda}_{(h)}}}{2\pi\hat \alpha_{\text{GUT}} }\bar u(p_3)v(p_4)
={\kappa_4M_{{\lambda}_{(h)}}}\bar u(p_3)v(p_4) {\alpha^{\prime}}_{\text{DM}}^\lambda \ .
\end{align}
As above, we have defined the parameters
\begin{equation}
\label{pen1}
-\frac{1}{2}{\alpha^{\prime}}^A_{\text{SM}}={\alpha^{\prime}}_{\text{SM}}^\lambda={\alpha^{\prime}}_{\text{DM}}^\lambda= \frac{\langle s \rangle}{2 \pi{\hat{\alpha}}_{GUT}} \ .
\end{equation} \ .

\subsubsection{Squared Amplitudes}

Putting everything together, we can now calculate the squared amplitudes for the dark matter production processes shown in Figures 1 and 2. Generically these are complex calculations with intricate and complicated results. However, as we will discuss in subsection 4.3 below, the dark matter production rate will become exponentially Boltzman suppressed whenever the center of mass energy $\sqrt{{\mathpzc s}}$ 
of the incoming observable particles in the thermal bath falls substantially below the mass of the dark matter particles, which we will ``generically'' denote by $m_{DM}$, and the masses of the $\phi^{2}$ and $\eta^{2}$ moduli\footnote{Note that we use $\mathpzc{s}$ to distinguish the center of mass-energy from the symbol $s$ used for the dilaton real scalar component.}. It is important to note that $\sqrt{{\mathpzc s}}\sim T$, where $T$ is the temperature of the thermal bath expressed as energy using the Boltzmann constant. Henceforth in this paper, for simplicity, we will { \it denote both the universal K\"ahler modulus and the temperature of the thermal bath by the same symbol ``T''. The meaning of the symbol will always be clear from the context}. We will therefore restrict our calculations to the case where
\begin{equation}
\label{pen2}
T^2 \gg m_{DM}^2,~m_{\phi^{2}}^2, ~m_{\eta^{2}}^2\ .
\end{equation}
The computation of the production processes then greatly simplifies. For the graphs in Figure 1 we find that
  \begin{align}
    |\mathcal{M}_{CC\rightarrow CC}|^2&\simeq\left[\alpha^C_{\text{SM}}\right]^2\left[\alpha^C_{\text{DM}}\right]^2\>\kappa_4^4{\mathpzc s}^2 \ ,\label{eq:CC-CC}\\
    |\mathcal{M}_{CC\rightarrow \lambda\lambda}|^2&\simeq\left[\alpha^C_{\text{SM}}\right]^2\left[\alpha^\lambda_{\text{DM}}\right]^2\>\kappa_4^4M^2_{\lambda_{(h)}}{\mathpzc s} \ ,\label{eq:CC-LL}\\
    |\mathcal{M}_{\psi \psi\rightarrow CC}|^2&\simeq 4\left[\alpha^\psi_{\text{SM}}\right]^2\left[\alpha^C_{\text{DM}}\right]^2\>\kappa_4^4 M_{\psi_{(o)}}^2 {\mathpzc s} \label{eq:PP-CC}\ ,\\
 |\mathcal{M}_{\psi \psi\rightarrow \lambda\lambda}|^2&\simeq4\left[\alpha^\psi_{\text{SM}}\right]^2\left[\alpha^\lambda_{\text{DM}}\right]^2\>\kappa_4^4 M_{\psi_{(o)}}^2 M^2_{\lambda_{(h)}} \label{eq:PP-LL}\ ,\\
  |\mathcal{M}_{AA\rightarrow CC}|^2&\simeq\left[\alpha^A_{\text{SM}}\right]^2\left[\alpha^C_{\text{DM}}\right]^2\>\kappa_4^4 {\mathpzc s} ^2 \ ,\label{eq:AA-CC}\\
    |\mathcal{M}_{AA\rightarrow \lambda \lambda}|^2&\simeq\left[\alpha^A_{\text{SM}}\right]^2\left[\alpha^\lambda_{\text{DM}}\right]^2\>\kappa_4^4 M^2_{\lambda_{(h)}} {\mathpzc s}  \ ,\label{eq:AA-LL}\\
    |\mathcal{M}_{\lambda \lambda\rightarrow CC}|^2&\simeq\left[\alpha^{\lambda}_{\text{SM}}\right]^2\left[\alpha^C_{\text{DM}}\right]^2\>\kappa_4^4 M^2_{\lambda_{(o)}} {\mathpzc s} \label{eq:LL-CC}\ ,\\
    |\mathcal{M}_{\lambda \lambda\rightarrow \lambda \lambda}|^2&\simeq \left[\alpha^\lambda_{\text{SM}}\right]^2\left[\alpha^\lambda_{\text{DM}}\right]^2\>\kappa_4^4 M^2_{\lambda_{(o)}}M^2_{\lambda_{(h)}}\label{eq:LL-LL}\ .
    \end{align}
Similarly, we can compute the graphs in Figure 2. As demonstrated above, $m_{\eta^{2}} \simeq m_{\phi^{2}}$  and since the vertex coefficients in Figure 2 are identical to the associated coefficients in Figure 1--up to a minus sign which vanishes in the squared amplitude--we find that the squared amplitudes of the $AA \rightarrow \lambda \lambda$ and $\lambda \lambda \rightarrow \lambda \lambda$ processes in Figure 2 are identical to the same processes in Figure 1. Therefore, including the $\eta^2$-mediated channel, the production rates of gauginos are simply doubled. This effect becomes important for the calculation of the expected dark matter relic density, which will be done in the next section.


It follows from \eqref{eq:coupling_def11} and \eqref{eq:coupling_def22} that the values of the $s$ and $t$ moduli VEVs, $\langle s\rangle $ and $\langle t \rangle$, and the $\hat \alpha_{\text{GUT}}$ parameter are central in determining the strength of the vertex amplitudes for the processes we have just outlined.
The values of these quantities, however, will depend strongly on the choice of heterotic M-theory vacuum and the associated mechanism for moduli stabilization and supersymmetry breaking.  For specificity, in the remainder of this subsection, we will compute the value of the parameters in both \eqref{eq:coupling_def11} and \eqref{eq:coupling_def22} explicitly within the framework
of the $B-L$ MSSM heterotic vacuum with the anomalous $U(1)$ hidden sector presented in \cite{Ashmore:2020ocb}. A brief discussion of the $B-L$ MSSM vacuum state is given in Appendix C. Here, we simply use those results relevant to the calculation at hand. To begin, note that this vacuum arises from a compactification on a Schoen threefold with $h^{1,1}=3$ and, hence, has three real K\"ahler moduli $a^{i}$, $i=1,2,3$. It is convenient to define the rescaled moduli
\begin{equation}
\label{sun1}
t^{i}=\hat {R}\frac{a^{i}}{V}
\end{equation}
where $\hat{R}$ is the length modulus of the fifth-dimensional interval and 
\begin{equation}
\label{sun2}
V=\frac{1}{6}d_{ijk}a^{i}a^{j}a^{k} \ .
\end{equation}
The $d_{ijk}$ coefficients are the intersection numbers of the Schoen threefold presented in \cite{Braun:2005nv,Ashmore:2020ocb}. In terms of the three complexified K\"ahler moduli 
\begin{equation}
\label{sun3}
T^{i}=t^{i}+i\chi \ ,
\end{equation}
the K\"ahler potential $K_{T}$ is given by
\begin{equation}
K_T=-\kappa_{4}^{-2}\ln\left(\frac{1}{48} d_{ijk}(T^i+\bar T^i)(T^j+\bar T^j)(T^k+\bar T^k)\right)\ .
\end{equation}
Inserting \eqref{sun3} and using \eqref{sun1} and \eqref{sun2}, it follows that 
\begin{equation}
\label{sun4}
K_T= -3\kappa_4^{-2}\ln \hat R\ .
\end{equation}
Comparing this to expression \eqref{burt6} for the K\"ahler potential $K_T$ written in terms of the universal modulus $T$, we learn that in the $B-L$ MSSM 
\begin{equation}
\label{hole1}
t = \frac{\hat R}{2}\ .
\end{equation}
At the same time, we can identify
\begin{equation}
\label{hole2}
s=V\ .
\end{equation}

Scanning over the physically viable points  in the ``magenta'' region of real K\"ahler moduli space--shown in Appendix C--and using the formalism presented in \cite{Ashmore:2020ocb} we find that  $\langle \hat{R}\rangle$  and $\langle V\rangle$ and must lie in the intervals
\begin{equation}
\label{park1}
\langle \hat R \rangle \in [1,5]\ ,
\end{equation}
and 
\begin{equation}
\label{park2}
\langle V\rangle \in [0.55,1.2]\ .
\end{equation}
Using the fact that
\begin{equation}
\label{park3}
\epsilon_{S}=\frac{\langle V \rangle^{1/3}}{\pi \langle \hat{R} \rangle} \ ,
\end{equation}
it follows from \eqref{rev5}, \eqref{hole1}, \eqref{hole2} and \eqref{park3} that, in this explicit example, the expression for $\beta$ simplifies to 
\begin{equation}
\label{easy1}
\beta=6 \langle V \rangle ^{2/3} \ .
\end{equation}
Hence, in the magenta region, the parameter $\beta$ must lie in the interval
\begin{equation}
\label{park5}
\beta \in [4.0, 6.8] \ .
\end{equation}
%
Therefore, the physically allowed values of the couplings defined in eq.\eqref{eq:coupling_def11} must have the approximate values
\begin{equation}
\label{eq:coupling_val1} 
\alpha^C_{\text{SM}}=\alpha_{\text{SM}}^\Psi=\alpha^C_{\text{DM}}=\frac{2}{\sqrt{3}} \frac{1} {(1+\frac{\beta^{2}}{3} )^{1/2}}  \in [0.28,0.46 ] \ .
\end{equation}
Similarly, by definition 
\begin{equation}
\hat\alpha_{\text{GUT}}=\langle \alpha_{u} \rangle V\ ,
\end{equation}
where 
$\langle \alpha_{u} \rangle $ is the MSSM unification coupling, chosen in the $B-L$ MSSM simultaneous Wilson line scenario to be
\begin{equation}
\langle \alpha_{u} \rangle =\frac{1}{26.64} \ ,
\end{equation}
and $V$ is the volume modulus defined in \eqref{sun2}.
Hence, the couplings defined in eq. \eqref{eq:coupling_def22} have the fixed value
\begin{equation}
\label{eq:coupling_val2} 
\tfrac{1}{2}\alpha^A_{\text{SM}}=\alpha_{\text{SM}}^\lambda=\alpha^\lambda_{\text{DM}}=\frac{1 }{2\pi \alpha_{u}} = 4.2 \ .
\end{equation}
The range of values we have obtained for the couplings in \eqref{eq:coupling_val1} and \eqref{eq:coupling_val2} will be important in our discussion of the observed relic density that this anomalous $U(1)$ hidden sector $B-L$ MSSM theory can produce.

\subsection{Dark Matter Production Mechanism--Non-Anomalous Case}


In Subsection 3.4, we identified three main moduli channels which couple the fields in the observable sector to the hidden sector fields--through the $\delta \tilde t$ modulus, the $\delta \tilde s$ modulus and the $\delta\tilde \sigma$ modulus. These moduli fields were defined in \eqref{eq:eigen_st} and \eqref{eq:eigen_st2}. As discussed above and in Appendix B, they are all massive of ${\cal{O}}(m_{SUSY})$. We have argued that when these moduli states are approximately unmixed after diagonalizing the moduli mass matrix, such that $\delta \tilde s\simeq \kappa_4^{-1}\delta s$,  $\delta \tilde t\simeq \kappa_4^{-1}\delta t$ and $\delta \tilde \sigma \simeq \kappa_4^{-1}\delta \sigma$, then we expect a separation in the interactions between the fields of the observable and the hidden sectors. Specifically, the $\delta \tilde t$ modulus couples the chiral matter multiplets from the observable and the hidden sectors only, while the $\delta \tilde s$ and $\delta \tilde \sigma$ moduli couple the gauge multiplets from the two sectors to each other.

{
\begin{figure}[t]
   \centering
       \begin{subfigure}[b]{0.436\textwidth}
       \centering
   \includegraphics[width=1\textwidth]{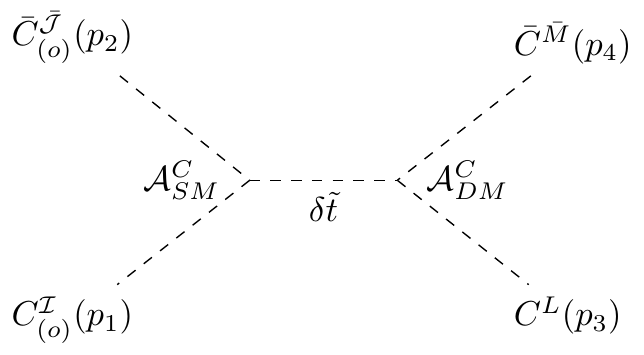}
\caption{}
\label{fig:int_1a}
\end{subfigure}\hfill
       \begin{subfigure}[b]{0.436\textwidth}
\centering
 \includegraphics[width=1\textwidth]{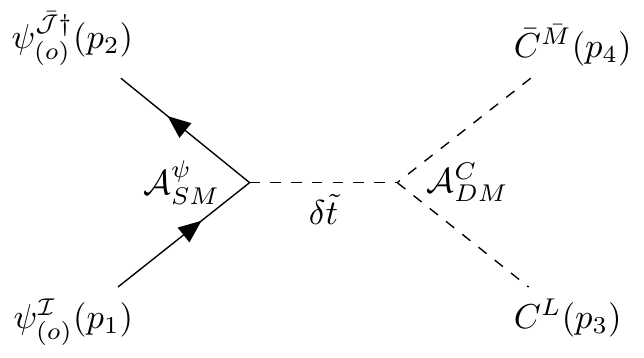}
\caption{}
\label{fig:int_1b}
\end{subfigure}
\caption{ Possible chiral dark matter production mechanisms. The massive $\delta \tilde t$ modulus can be produced by scalar-scalar and fermion-fermion scattering in the thermal bath of particles in the observable sector created during reheating. Based on the expected mass spectrum shown in Table 1 and the possible interactions between the observable sector thermal bath and the hidden sector fields, we identified the hidden matter scalars $C_{(h)}^L$, $L=1,\dots, {N}_{(h)}$, as one of the plausible dark matter candidates within our theoretical context. }
\label{fig:int_1}
\end{figure}
\begin{figure}[t]
   \centering
       \begin{subfigure}[b]{0.436\textwidth}
       \centering
   \includegraphics[width=1\textwidth]{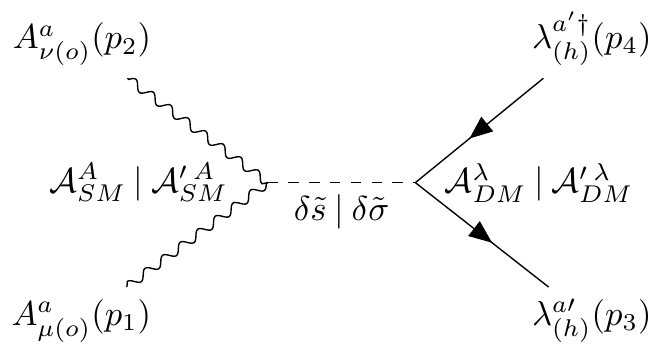}
\caption{}
\label{fig:int_2a}
\end{subfigure}\hfill
       \begin{subfigure}[b]{0.436\textwidth}
\centering
 \includegraphics[width=1\textwidth]{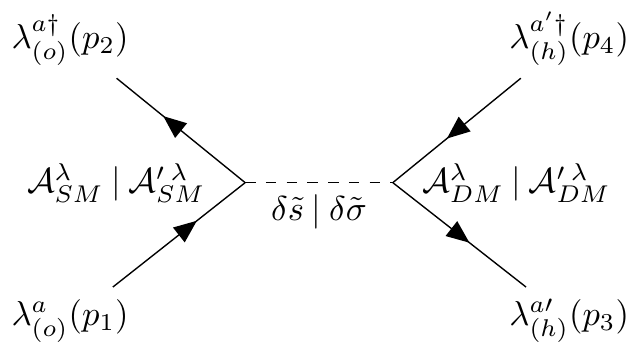}
\caption{}
\label{fig:int_2b}
\end{subfigure}
\caption{ Possible gaugino dark matter production mechanisms. The massive $\delta \tilde s$ modulus can be produced during the gauge boson-gauge boson and gaugino-gaugino scattering in the thermal bath of particles in the observable sector during reheating. Based on the expected mass spectrum shown in Table 1 of Appendix B, and the possible interactions between the observable sector thermal bath and the hidden sector fields, we identified the hidden sector gauginos $\lambda_{(h)}^{a^\prime}$ as one of the plausible dark matter candidates within our theoretical context.  For the observable sector gauginos, the index $a$ runs over the adjoint representation of $SU(3)_{C} \times SU(2)_{L} \times U(1)_{Y}$ and any further extension of the standard model gauge group--such as $U(1)_{B-L}$ in the $B-L$ MSSM. For the hidden sector gauginos, the index $a^{\prime}$ runs over the adjoint representation of the hidden sector group $\mathcal{H}_1$ which does not
contribute to gaugino condensation.}
\label{fig:int_2}
\end{figure}
}

\subsubsection{$\delta \tilde t$ modulus portal:}

The coupling of the $\delta \tilde t$ modulus to the ${N}_{(h)}$ matter fermions $\psi^L_{(h)}$ in the hidden sector is proportional to the fermion mass. However, because these hidden sector fermions are {\it massless,} it is impossible to produce them--thus leaving {\it the hidden sector matter scalars as the only possible $\delta \tilde t$ decay product}.
The massive $\delta \tilde t$ modulus can be produced by scalar-scalar and fermion-fermion scattering in the thermal bath of observable sector particles during reheating. The processes producing dark matter composed of the hidden scalars $C_{(h)}^L$ are displayed in Figure 3. The strength of interaction vertices on the observable sector was shown in subsection 3.4 to be
\begin{equation}
\mathcal{A}_{SM}^C=\delta_{{\cal I}\bar {\cal J}}\>p_{1\mu}p_2^{\mu}\>\frac{\kappa_4}{\langle t\rangle}\equiv 
\delta_{{\cal I}\bar {\cal J}}\>p_{1\mu}p_2^{\mu}\>\kappa_4\alpha^C_{\text{SM}}
\end{equation}
for the scalar components and 
\begin{equation}
\mathcal{A}_{SM}^\psi=\delta_{{\cal I}\bar {\cal J}}\>\frac{\kappa_4M_{\Psi^{\cal I}_{(o)}}}{\langle t\rangle }\bar u(p_1)v(p_2) 
=\delta_{{\cal I}\bar {\cal J}}\>{\kappa_4M_{\Psi^{\cal I}_{(o)}}}\bar u(p_1)v(p_2) \alpha_{\text{SM}}^\psi 
\end{equation}
for the fermions, while the coupling on the hidden sector was found in subsection 3.4 to be
\begin{equation}
\mathcal{A}^C_{DM}=\delta_{L\bar M}\>p_{3\mu}p_4^{\mu}\>\frac{\kappa_4}{\langle t\rangle}\equiv \delta_{L\bar M}\>p_{3\mu}p_4^{\mu}\>{\kappa_4}\alpha^C_{\text{DM}}\ ,
\end{equation}
where
\begin{equation}
\label{eq:coupling_def1}
\alpha^C_{\text{SM}}=\alpha_{\text{SM}}^\Psi=\alpha^C_{\text{DM}}=\frac{1}{\langle t\rangle}\ .
\end{equation}

\subsubsection{$\delta \tilde s$ and $\delta \tilde \sigma$ moduli portals:}

On the other hand, the $\delta \tilde s$ and $\delta \sigma$ moduli can decay into both the massive hidden sector gauginos $\lambda_{(h)}^{a^{\prime}}$ and the massless hidden sector gauge bosons $A_{(h)\mu}^{a^{\prime}}$.
Of the two, since the mass of the gauge fields vanish, {\it only the gauginos can be a dark matter $\delta \tilde s$ decay product}. The $\delta \tilde s$ modulus can be produced during the boson-boson and gaugino-gaugino scattering in the thermal bath of observable sector particles during reheating. The processes of producing possible gaugino dark matter are displayed in Figure 4. The strength of the interaction vertices on the observable sector is shown in subsection 3.4 to be
\begin{equation}
\mathcal{A}_{SM}^A=-\mathcal{A}_{SM}^{\prime A}=p_{1\mu}p_2^{\mu}\>\frac{\kappa_4}{\pi\hat \alpha_{\text{GUT}}}\equiv 
p_{1\mu}p_2^{\mu}\>\kappa_4\alpha^A_{\text{SM}}
\end{equation}
for the vector components and 
\begin{equation}
\mathcal{A}_{SM}^\lambda=\mathcal{A}_{SM}^{\prime \lambda}=\frac{\kappa_4M_{{\lambda}_{(o)}}}{2\pi\hat \alpha_{\text{GUT}} }\bar u(p_1)v(p_2) 
={\kappa_4M_{{\lambda}_{(o)}}}\bar u(p_1)v(p_2) \alpha_{\text{SM}}^\lambda
\end{equation}
for the gauginos, while the coupling to the hidden sector gauginos was found to be
\begin{equation}
\mathcal{A}^\lambda_{DM}=\mathcal{A}^{\prime \lambda}_{DM}
=\frac{ \kappa_4 M_{\lambda_{(h)}}}{2\pi\hat \alpha_{\text{GUT}} }\bar u(p_3)v(p_4) 
=\kappa_4 M_{\lambda_{(h)}} \bar u(p_3)v(p_4) \alpha_{\text{DM}}^\lambda\ ,
\end{equation}
where
\begin{equation}
\label{eq:coupling_def2}
\tfrac{1}{2}\alpha^A_{\text{SM}}=\alpha_{\text{SM}}^\lambda=\alpha^\lambda_{\text{DM}}=\frac{1}{2\pi\hat \alpha_{\text{GUT}}}\ .
\end{equation}

\subsubsection{Squared Amplitudes}

Exactly as in the anomalous $U(1)$ case discussed above, in the limit that 
\begin{equation}
\label{pencil1}
T^2 \gg m_{DM}^2,~m_{\tilde{t}}^2, ~m_{\tilde{s}}^2, ~m_{\tilde{\sigma}}^2 \ ,
\end{equation}
the squared amplitudes associated with the four processes shown in Figures \ref{fig:int_1} and \ref{fig:int_2} are given by
    \begin{align}
    |\mathcal{M}_{CC\rightarrow CC}|^2&\simeq\left[\alpha^C_{\text{SM}}\right]^2\left[\alpha^C_{\text{DM}}\right]^2\>\kappa_4^4{\mathpzc s}^2 \ ,\label{eq:CC-CC}\\
    |\mathcal{M}_{\psi \psi\rightarrow CC}|^2&\simeq4\left[\alpha^\psi_{\text{SM}}\right]^2\left[\alpha^C_{\text{DM}}\right]^2\>\kappa_4^4 M_{\psi_{(o)}}^2 {\mathpzc s} \label{eq:PP-CC}\ ,\\
    |\mathcal{M}_{AA\rightarrow \lambda \lambda}|^2&\simeq 2\left[\alpha^A_{\text{SM}}\right]^2\left[\alpha^\lambda_{\text{DM}}\right]^2\>\kappa_4^4 M^2_{\lambda_{(h)}} {\mathpzc s}  \ ,\label{eq:AA-LL}\\
    |\mathcal{M}_{\lambda \lambda\rightarrow \lambda \lambda}|^2&\simeq 2\left[\alpha^\lambda_{\text{SM}}\right]^2\left[\alpha^\lambda_{\text{DM}}\right]^2\>\kappa_4^4 M^2_{\lambda_{(o)}}M^2_{\lambda_{(h)}}\label{eq:LL-LL}\ ,
    \end{align}
where, again, $\sqrt{{\mathpzc s}} \sim T$ is the center of mass energy of the two incoming particles from the observable sector thermal bath. The squared amplitudes of the processes $AA\rightarrow \lambda \lambda$ and $\lambda \lambda\rightarrow \lambda \lambda$ were calculated by summing over both the $\delta \tilde s$ and $\delta \tilde \sigma$ channels.

\subsection{Dark Matter Relic Density}

As discussed in Appendix A, following the inflationary epoch the inflaton field $\psi$ begins a period of coherent oscillations with a potential energy of the form
\begin{equation}
\label{fin1}
V (\psi)\simeq \frac{1}{2}m^{2}\psi^{2} \ ,
\end{equation}
during which it begins to decay into the particles of the supersymmetric observable sector. The size of the inflaton field at the beginning of this oscillatory period is simply denoted by $\psi_{osc}$ and the associated potential energy by $V_{osc}$. 
%
%
During this oscillatory period, called the `` reheating'' or ``inflation dominated (ID)'' epoch, more and more supersymmetric particles are produced and the potential energy of the inflaton rapidly decreases. At some point, all particles in the bath achieve thermal equilibrium and the inflaton potential goes to zero. This occurs at the reheating temperature $T_{RH}$. Be that as it may, it is useful to define the ``maximum temperature'' attained during reheating as \cite{Kolb:1990vq}
\begin{equation}
\label{fin3}
T_{max}=(V_{osc}^{1/4}T_{RH})^{1/2} \ .
\end{equation}
After thermal equilibrium is achieved at $T_{RH}$, the Universe enters the so-called ``radiation dominated (RD)'' epoch, where the temperature continues to decrease as the universe expands.
That is--after inflation--as the Universe cools down, the temperature transitions as
\begin{align}
&\text{ID}: \quad { T}_{max} > T > {T}_{RH} \ ,\\
&\text{RD}: \quad {T}_{RH} > T >{T}_{0} \ ,
\end{align}
where $T_{0}$ is the temperature of the present Universe.

The dark matter relic density, see for example \cite{Chowdhury:2018tzw}, is given by
\begin{equation}
\label{train1}
\Omega h^2=\frac{m_{\text{DM}}n_{\text{DM}}}{\rho_c }\ ,
\end{equation}
 where $m_{\rm{DM}}$ is the ``generic'' dark matter mass, $n_{\rm{DM}}$ is the dark matter number density and  $\rho_c$ the critical density today, can be produced during both the inflaton dominated and the radiation dominated eras. Following~\cite{Chowdhury:2018tzw}, the expression for the dark matter relic density observed today is
\begin{equation}
\begin{split}
\label{eq:OmegaDM}
\Omega h^2&\simeq \Omega h^2_{\rm{ID}}+\Omega h^2_{{\rm{RD}}}\\
&\approx 4\times 10^{24}\>m_{\text{DM}}\left(  1.07\times  {T}_{\text{RH}}^7\>\int_{{ T}_{\text{RH}}}^{{T}_{\text{max}}} 
d {T}\frac{R( { T})}{ { T}^{13}}+\int_{  T_{0}}^{ {T}_{\text{RH}}}d { T}\frac{R( { T})}{ { T}^6}  \right)\ ,
\end{split}
\end{equation}
where $R(T)$ is the dark matter production rate at temperature $T$.
When the temperature of the Universe drops below the dark matter particle mass, the dark matter production rate becomes exponentially suppressed by the Boltzmann factor
    \begin{equation}
    \label{bol}
    R( {T})\sim e^{-m_{\text{DM}}/ {T}}\ . 
    \end{equation}
 Effectively, therefore, dark matter production stops in the regime where
    \begin{equation}
    \label{fin4}
  {T}<m_{\text{DM}}  \Rightarrow  R( {T})\approx 0\ .
    \end{equation}

    As discussed in Appendix A, in our inflation model it is necessary for the scale of $N=1$ supersymmetry breaking to be 
\begin{equation}
\label{pencil2}
m_{SUSY} \sim {\cal{O}}(10^{13}~{\rm GeV}) .
\end{equation}
Further analysis showed that the inflaton oscillations start, and the reheating temperature is given by,
\begin{equation}
\label{pencil3}
T_{max}=1.623\times 10^{14} {\rm GeV}  \quad {\rm and}  \quad T_{RH}=1.13 \times 10^{13} {\rm GeV}
\end{equation}
respectively.  It follows from our discussion in Section 3.1 that when the hidden sector contains an anomalous $U(1)$ subgroup then when $N=1$ supersymmetry is broken at scale \eqref{pencil2}, the ``generic'' dark matter mass and the moduli masses are given by
\begin{equation}
\label{pol1}
m_{DM} \simeq m_{\phi^{2}} \simeq m_{\eta^{2}} \sim {\cal{O}}(10^{13}~{\rm GeV}) \ .
\end{equation}
Similarly, for a non-anomalous hidden sector, it follows from subsection 4.2 and Table 1 in Appendix B that
\begin{equation}
\label{pol2}
m_{DM} \simeq m_{\tilde{t}} \simeq m_{\tilde{s}} \simeq m_{\tilde{\sigma}}  \sim {\cal{O}}(10^{13}~{\rm GeV}) \ .
\end{equation}
We see, therefore, from \eqref{pen2} and \eqref{pencil1} that in all cases, when supersymmetry is broken as in \eqref{pencil2}, the dominant contributions to the squared dark matter production amplitudes are for temperatures $T$ contained in the inflaton-dominated regime--that is, for 
\begin{equation}
\label{pol3}
T \in [T_{max}, T_{RH}] \ ,
\end{equation}
where $T_{max}$ and $T_{RH}$ are given in \eqref{pencil3}.
In summary, although we cannot determine the exact masses of the moduli mediators or the hidden sector matter scalars and gauginos, which play the role of dark matter, without giving an explicit model of supersymmetry breaking,  we do expect that they have values close to the supersymmetry breaking scale, $m_{\text{SUSY}}\approx 10^{13}$GeV--which is about the same size as the reheating temperature. Given this expected mass hierarchy within our context, it becomes clear that dark matter production becomes heavily suppressed when the temperature of the universe drops below $ T_{RH}$. It follows that most of the hidden sector matter must be produced during the inflaton-dominated era, as the Universe cools down from $T_{max}$ to $T_{RH}$. 

   The squared amplitudes of the processes involving fermions--in both the anomalous and non-anomalous hidden sector scenarios--are proportional to the squares of the fermion masses $M_{\psi_{(o)}}$, $M_{\lambda_{(o)}}$ or $M_{\lambda_{(h)}}$. While the gauginos $\lambda_{(o)}^a$ and $\lambda_{(h)}^{a^{\prime}}$ become massive as a direct consequence of $N=1$ supersymmetry breaking, which sets the expected value of their masses to be around $m_{\text{SUSY}}\sim 10^{13}$GeV, the mechanism in which the $\mathcal{N}_{(o)}$ chiral fermions from the observable sector become massive is less straightforward. 
 As discussed in Appendix A, the mass of an observable sector fermion $\psi^{\cal I}_{(o)}$ is proportional to the root mean squared (r.m.s) value of the inflation during the reheating period,
   \begin{equation}
M_{\psi_{C(o)}^{\cal I}}=y_{\psi \psi^{\cal I}\psi^{\cal I}}\sqrt{\langle \psi \rangle^{2}}\ ,
\end{equation}
which is not constant as $ T$ drops from $ T_{\max}$ to $ {T}_{RH}$.
A key insight regarding the $\mathcal{N}_{(o)}$ fermions $\psi_{(o)}^{\cal I}$ is that irrespective of what the value of the r.m.s. of the inflaton is at any given 
$ {T}\in[ {T}_{RH},  {T}_{max}]$, they cannot be part of the thermal bath unless their production is kinematically allowed. 
It is shown in~\cite{Cai:2018ljy} that these fermions can be pair-produced in inflaton decays of the type $\psi\rightarrow \psi^{\cal I}_{(o)}\psi^{\cal{I}\dag}_{(o)}$ only if their mass $M_{\psi^{\cal{I}}_{(o)}}$ is less than half of the inflaton mass; that is
\begin{equation}
M_{\psi^{\cal{I}}_{(o)}}\leq m_{\psi}/2=0.79\times 10^{13}\text{GeV}\ .
\end{equation}
As we will now show, the fact that in both the anomalous and non-anomalous hidden sector cases
\begin{equation}
\label{fd1}
M_{\psi_{(o)}}, M_{\lambda_{(o)}}, M_{\lambda_{(h)}} \lesssim m_{\text{SUSY}}\sim 10^{13} {\rm{GeV}} \ ,
\end{equation}
implies that producing dark matter through any process involving either an observable or hidden sector fermion
is subdominant compared to only producing it through processes involving bosons. 
%

Let us first consider the squared amplitude decay processes in the anomalous hidden sector case presented in subsection 4.1. In this scenario, there are only two decay processes that do not involve any fermions, specifically $CC \rightarrow CC$ and $AA \rightarrow CC$ as shown in Figure 1. To begin, we consider the ratio of the squared amplitudes for any process involving  a fermion to $CC \rightarrow CC$. We will return to the $AA \rightarrow CC$ process later. First consider the $\psi \psi \rightarrow CC$ process shown in Figure 1. Using the the results in subsection 4.1, we find that
\begin{equation}
\label{fd2}
\frac{  |\mathcal{M}_{\psi \psi\rightarrow CC}|^2}{ |\mathcal{M}_{CC\rightarrow CC}|^2}\sim \frac{M_{\psi}^2}{{\mathpzc s}}\sim
\frac{M_{\psi}^2}{ { T}^2}\ .
\end{equation}
Using \eqref{fd1} and the fact that ${T}\in [ {T}_{max},\> { T}_{RH}]$, it follows that the ratio
\begin{equation}
\label{fd3}
\frac{M_{\psi}^2}{ { T}^2} \ll 1
\end{equation}
and, hence, dark matter produced in the process $\psi\psi\rightarrow CC$ is negligible compared to dark matter produced in $CC\rightarrow CC$, for any observable sector fermions and any scalars $C$. Similarly, it follows from \eqref{fd1} and the fact that ${T}\in [ {T}_{max},\> { T}_{RH}]$ that any processes involving either observable or hidden sector gauginos, or both, are
also suppressed relative to $CC \rightarrow CC$. For example,
\begin{equation}
\begin{split}
\frac{  |\mathcal{M}_{AA\rightarrow \lambda \lambda}|^2}{ |\mathcal{M}_{CC\rightarrow CC}|^2}&\sim \frac{M_{\lambda^{a^{\prime}}_{(h)}}^2}{{\mathpzc s}}\sim
\frac{M_{\lambda^{a^{\prime}}_{(h)}}^2}{ {T}^2}\ll 1\ ,\label{gaugasf1}\\
\frac{  |\mathcal{M}_{\lambda\lambda \rightarrow \lambda \lambda}|^2}{ |\mathcal{M}_{CC\rightarrow CC}|^2}&\sim \frac{M_{\lambda^a_{(o)}}^2M_{\lambda^{a^{\prime}}_{(h)}}^2}{{\mathpzc s}^2}\sim
\frac{M_{\lambda^a_{(o)}}^2M_{\lambda^{a^{\prime}}_{(h)}}^2}{ {T}^4}\ll1\ .
\end{split}
\end{equation}
To complete the anomalous hidden sector case, however, we must consider the relative strengths of the $CC \rightarrow CC$ and $AA \rightarrow CC$ pure scalar processes. Using the results in subsection 4.1, we find that
\begin{equation}
\label{fd4}
\frac{|\mathcal{M}_{CC \rightarrow CC}|^2}{ |\mathcal{M}_{AA \rightarrow CC}|^2} \sim \frac{\left[\alpha^{C}_{\text{SM}}\right]^2}{\left[\alpha^{A}_{\text{SM}}\right]^2} \ .
\end{equation}
It then follows from \eqref{eq:coupling_val1} and \eqref{eq:coupling_val2} that, for all K\"ahler moduli in the magenta region, the ratio 
\begin{equation}
\label{fd5}
\frac{\left[\alpha^{C}_{\text{SM}}\right]^2}{\left[\alpha^{A}_{\text{SM}}\right]^2}\ll 1 \ .
\end{equation}
We conclude, therefore, that {\it $AA \rightarrow CC$ is the dominant process to produce dark matter in the anomalous hidden sector case}. 

Let us now consider the case of a non-anomalous hidden sector, as presented in subsection 4.2. In this scenario, there is only one dark matter production process that does not involve any fermions, specifically $CC \rightarrow CC$ in Figure 3.  As in the anomalous hidden sector case, the production amplitudes for the three other processes that do involve fermions--see the squared amplitudes in subsection 4.2--are subdominant with respect to the pure scalar case. For example, for the $\psi\psi \longrightarrow CC$ process in Figure 3, we find that 
\begin{equation}
\label{nora1}
\frac{  |\mathcal{M}_{\psi \psi\rightarrow CC}|^2}{ |\mathcal{M}_{CC\rightarrow CC}|^2}\sim \frac{M_{\psi}^2}{{\mathpzc s}}\sim
\frac{M_{\psi}^2}{ { T}^2} \ll 1
\end{equation}
where we have used \eqref{fd1} and the fact that ${T}\in [ {T}_{max},\> { T}_{RH}]$. Similarly, we find that 
\begin{equation}
\begin{split}
\frac{  |\mathcal{M}_{AA\rightarrow \lambda \lambda}|^2}{ |\mathcal{M}_{CC\rightarrow CC}|^2}&\sim \frac{M_{\lambda^{a^{\prime}}_{(h)}}^2}{{\mathpzc s}}\sim
\frac{M_{\lambda^{a^{\prime}}_{(h)}}^2}{ {T}^2}\ll 1\ ,\label{gaugasf1}\\
\frac{  |\mathcal{M}_{\lambda\lambda \rightarrow \lambda \lambda}|^2}{ |\mathcal{M}_{CC\rightarrow CC}|^2}&\sim \frac{M_{\lambda^a_{(o)}}^2M_{\lambda^{a^{\prime}}_{(h)}}^2}{{\mathpzc s}^2}\sim
\frac{M_{\lambda^a_{(o)}}^2M_{\lambda^{a^{\prime}}_{(h)}}^2}{ {T}^4}\ll1\ .
\end{split}
\end{equation}
It follows that, unlike the anomalous case, for a { \it non-anomalous hidden sector the dominant dark matter production mechanism is the $CC \rightarrow CC$ process} given in Figure 3.

\subsection {Dark Matter Production Rate: Anomalous Case} 

   The rate of production $R( T)$ of scalar hidden sector matter via the dominant mechanism of the type $AA\rightarrow CC$, shown in Figure 1, has been computed in~\cite{Chowdhury:2018tzw}. To do this, it is first necessary to recognize that not all hidden sector scalars $C_{(h)}^{L}, L=1\dots,N_{(h)}$ necessarily have the same mass--see, for example, expression  \eqref{end2} and the associated discussion in Appendix C. The contribution of those scalars with mass substantially larger that $m_{DM}$ will, using \eqref{bol}, be strongly Boltzmann suppressed. Furthermore, the scalars with mass $m \ll m_{DM}$ will have their contributions suppressed by the ratio $m / m_{DM}$. It follows that only those ${\hat{N}}_{(h)} \leq N_{(h)}$ hidden sector scalars with masses $m \approx m_{DM}$ can substantially contribute to $R(T)$. It was shown in \cite{Chowdhury:2018tzw} that the rate $R(T)$ to produce $\hat{N}_{(h)}$ dark matter scalars of mass $m \approx m_{DM}$ is given by
\begin{equation}
\label{tia1}
R(T)=\hat{N}_{(h)} \frac{\pi^3}{21,600}\>{[\alpha^A_{\text{SM}}]}^2{[\alpha^C_{\text{DM}}]}^2\>\kappa_4^4 {T}^8\ .
\end{equation}
Evaluating expression \eqref{eq:OmegaDM} by integrating $T$ in the reheating regime between $ {T}_{max}$ and $3.33 ~{T}_{RH}$--where the required limit ${T}^2\gg m^2_{SUSY}$ is satisfied--and using the heavy regime production rate $R( {T})$ given in eq. \eqref{tia1}, we find that
\begin{equation}
\begin{split}
\label{eq:OmegaDM3}
\Omega h^2&\approx \Omega h^2_{\text{ID}}\\
&\approx (4.28\times 10^{24} ) \hat{N}_{(h)}\>\frac{\pi^3}{21,600}\>{[\alpha^A_{\text{SM}}]}^2{[\alpha^C_{\text{DM}}]}^2\>m_{DM}\>\kappa_4^4  {T}_{RH}^7\>\int_{3.33 T_{RH}}^{ {T}_{max}} d {T}\frac{1}{ {T}^{5}} \\
&\approx (8.7\times 10^{21}) \hat{N}_{(h)} \>\frac{\pi^3}{21,600}\>{[\alpha^A_{\text{SM}}]}^2{[\alpha^C_{\text{DM}}]}^2\>m_{DM}\>\kappa_4^4  {T}_{RH}^3\ .
\end{split}
\end{equation}
Note from \eqref{pencil3} that since $T_{max} \approx 10 ~T_{RH}$, it is subdominant in the integral and can be ignored. Using $ {T}_{RH}=1.13 \times 10^{13}$ GeV, we obtain the observed value for the dark matter relic density, given, for example, in \cite{Planck:2018vyg} to be
\begin{equation}
\label{bird1}
\Omega h^2=0.12, 
\end{equation}
if
\begin{equation}
 \hat{N}_{(h)}{[\alpha^A_{\text{SM}}]}^2{[\alpha^C_{\text{DM}}]}^2\>m_{DM}\approx 3.4\times 10^{14}\text{GeV}\ .
\end{equation}
This can indeed be achieved for the dimensionless couplings $\alpha^v_{\text{SM}}, \alpha^C_{\text{DM}}$ in the ranges presented in \eqref{eq:coupling_val1}, \eqref{eq:coupling_val2} if the number of hidden sector  dark matter scalar components $ \hat{N}_{(h)}\sim \mathcal{O}(1-14)$ and if $m_{DM} \sim {\cal{O}}(m_{SUSY})$, which is, indeed, its natural scale. 

\subsection {Dark Matter Production Rate: Non-Anomalous Case} 

    The rate of production $R( {T})$ for ${\cal{N}}_{(0)}$ observable sector scalars to produce $\hat{N}_{(h)} \leq N_{(h)}$ hidden sector scalar fields via the mechanism of the type $CC\rightarrow CC$ shown in Figure~\ref{fig:int_1a}, has been computed in~\cite{Chowdhury:2018tzw}. In the limit that $m_{\tilde t}, m_{DM}\ll  {T}$, one finds 
\begin{equation}
\label{eq:RTHeavy}
R( T)={\cal{N}}_{(0)} \hat{N}_{(h)}\frac{\pi^3}{108,000}\>{[\alpha^C_{\text{SM}]}}^2{[\alpha^C_{\text{DM}}]}^2\>\kappa_4^4 {T}^8\ .
\end{equation}
Evaluating expression \eqref{eq:OmegaDM} by integrating $T$ in the reheating regime between $ {T}_{max}$ and $3.33~{T}_{RH}$--where the required limit ${T}^2\gg m^2_{SUSY}$ is satisfied--and using the heavy regime production rate $R( {T})$ given in eq. \eqref{eq:RTHeavy}, we find that
%
 \begin{equation}
\begin{split}
\label{eq:OmegaDM2}
\Omega h^2&\approx \Omega h^2_{\text{ID}}\\
&\approx (4.28\times 10^{24}) {\cal{N}}_{(o)}{\hat{N}}_{(h)}\>\frac{\pi^3}{108,000}\>{[\alpha^C_{\text{SM}}]}^2{[\alpha^C_{\text{DM}}]}^2\>m_{DM}\>\kappa_4^4  {T}_{\text{RH}}^7\>\int_{3.33 {T}_{RH}}^{ {T}_{max}} 
d {T}\frac{1}{ {T}^{5}} \\
&\approx (8.7\times 10^{21})  {\cal{N}}_{(o)}{\hat{N}}_{(h)}\>\frac{\pi^3}{108,000}\>{[\alpha^C_{\text{SM}}]}^2{[\alpha^C_{\text{DM}}]}^2\>m_{DM}\>\kappa_4^4  {T}_{RH}^3\ ,
\end{split}
\end{equation}
Again, noting from \eqref{pencil3} that since $T_{max} \approx 10 ~T_{RH}$ it is subdominant in the integral and can be ignored. Using $ {T}_{RH}=1.13 \times 10^{13}$ GeV, we obtain the observed dark matter relic density, $\Omega h^2=0.12$, for
%
\begin{equation}
 {\cal{N}}_{(o)}{\hat{N}}_{(h)}{[\alpha^C_{\text{SM}}]}^2{[\alpha^C_{\text{DM}}]}^2\>m_{DM}\approx 1.7\times 10^{15}\text{GeV}\ .
\end{equation}
This can be achieved if the dimensionless couplings $\alpha^C_{\text{SM}}, \alpha^C_{\text{DM}}\sim \mathcal{O}(1)$, if the number of scalar fields satisfy $ {\cal{N}}_{(o)}{\hat{N}}_{(h)}\sim \mathcal{O}(10-100)$, and if the dark matter mass $m_{DM}$ is around the SUSY breaking scale, which is, indeed, its natural scale. 

\subsubsection*{Acknowledgements}

We would like to thank Mark Trodden for helpful conversations. Sebastian Dumitru is supported in part by research grant DOE No.~DESC0007901. Burt Ovrut is supported in part by both the research grant DOE No.~DESC0007901 and SAS Account 020-0188-2-010202-6603-0338.\\

\appendix

\section{Observable Sector Higgs-Sneutrino Inflation}

The observable sector of the $B-L$ MSSM heterotic theory has the exact $N=1$ supersymmetric particle content as the MSSM with the addition of three right-handed neutrino chiral multiplets--one for each of the three families. The gauge group is that of the standard model, with an extra gauged $U(1)_{B-L}$ factor. Although the $B-L$ MSSM arises from compactification of heterotic M-theory on a Calabi-Yau threefold with $h^{1,1}=3$ and, hence, has three K\"ahler moduli, Higgs-Sneutrino inflation is developed by considering only the ``universal'' modulus. That is, the moduli of the cosmological theory are the dilaton $S$ and a single K\"ahler modulus $T$. The K\"ahler potential of the observable sector, which arises by restricting the full $h^{1,1}=3$ theory to the universal modulus, is given by 
\begin{equation}
\label{A1}
K=-\kappa_4^{-2}\ln(S+\bar S)-3\kappa_4^{-2}\ln\left(T+\bar T- \kappa_{4}^{2}\mathcal{G}_{{\cal{I}}{\bar{\cal{J}}} } C_{(o)}^{\cal{I}}\bar{C}_{(o)}^{\bar {\cal{J}}}\right) \ ,
\end{equation}
where $C_{(o)}^{\cal{I}}$ are the dimension one scalars of the $B-L$ MSSM observable sector. $\mathcal{G}_{{\cal{I}}{\bar{\cal{J}}}}$ are generically complex structure dependent hermitian matrices--which were chosen, for simplicity, to be $\delta_{{\cal{I}}{\bar{\cal{J}}}}$.

It was shown in \cite{Deen:2016zfr,Cai:2018ljy} that a linear combination of the up, neutral Higgs scalar $H_{u}^{0}$, and the left-handed and right-handed sneutrinos, $\nu_{L,3}$ and $\nu_{R,3}$ respectively, produce a viable candidate for an inflationary scalar.  In this Appendix, we provide a brief outline of this Higgs-Sneutrino inflation theory--along with a discussion of several of its properties relevant to this paper.

Specifically, the inflaton is given by the real part of 
\begin{equation}
\label{A2}
\phi_{1}=\frac{1}{\sqrt{3}} H_{u}^{0}+\nu_{L,3}+\nu_{R,3} \ .
\end{equation}
It is convenient when discussing the Lagrangian for $\phi_{1}$ to work in units with $\kappa_{4}^{-1}$ set to unity--that is, setting the reduced Planck mass $M_{P}/\sqrt{8 \pi}=1$. Then, to canonically normalize the kinetic energy, one defines a real scalar field $\psi$ by
\begin{equation}
\label{A3}
\phi_{1}=\sqrt{3} \tanh \left( \frac{\psi}{\sqrt{6}}\right) \ .
\end{equation}
The relevant part of the $\psi$ potential energy is given by the soft supersymmetry breaking potential
\begin{equation}
\label{A4}
V_(\psi)=3 m^{2} \tanh^{2} \left( \frac{\psi}{\sqrt{6}} \right) \ . 
\end{equation}
where
\begin{equation}
\label{A5}
m^{2}=\frac{1}{3} ( m^{2}_{H_{u}^{(0)}}+ m^{2}_{\nu_{L,3}}+ m^{2}_{\nu_{R,3}}) \ .
\end{equation}
To satisfy the $Planck ~2015$ cosmological data, the soft mass parameter $m$ must, after restoring the Planck units to GeV, take the value
\begin{equation}
\label{A6}
m=1.58 \times 10^{13}~GeV  \ .
\end{equation}
Hence, supersymmetry in this inflationary $B-L$ MSSM must be broken at a high scale 
\begin{equation}
\label{A7}
m_{SUSY} \sim {\mathcal{O}}(10^{13}~ \rm{GeV}) \ .
\end{equation}
It then follows that the Universe acquires 60 e-foldings of inflation as the inflaton rolls from $\psi_{*} \simeq 6.23$, where $V_{*}^{1/4} \simeq 7.762 \times 10^{15}~\rm{GeV}$,  to $\psi_{end} \simeq 1.21$, where $V_{end}^{1/4} \simeq 5.140 \times10^{15} \rm{GeV}$.

As shown in Figure 5, at the end of inflation, the inflaton enters an oscillatory phase with potential
\begin{equation}
\label{A8}
V(\psi) \simeq \frac{1}{2} m^{2}\psi^{2} \ .
\end{equation}
\begin{figure}[t]
   \centering
       \begin{subfigure}[b]{0.76\textwidth}
       \centering
   \includegraphics[width=1\textwidth]{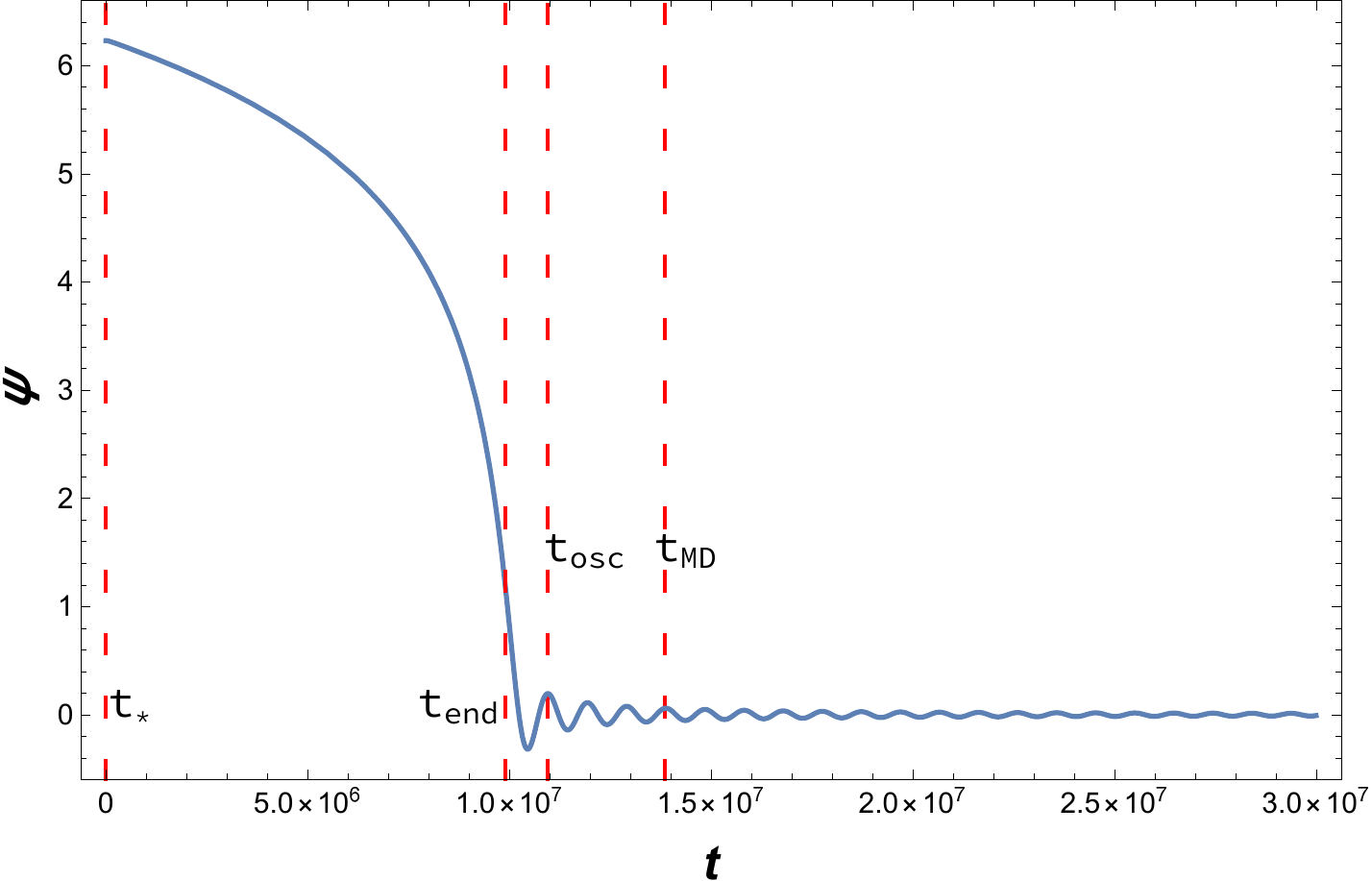}
\end{subfigure}
\caption{
The numerical solutions for $\psi(t)$ and $H(t)$, where we have set $\kappa_{4}^{-1}=1$. Note that $t_*=0$ and $t_{end}\simeq 9.89\times 10^6$ mark the beginning and end of the inflationary period.  The times $t>t_{end}$ correspond to the post inflationary epoch. The time $t_{osc}\simeq 1.096\times 10^7$ marks the point at which the potential energy is well approximated by $V=\frac{1}{2} m^{2}{\psi^{2}}$ and $t_{MD}\simeq 1.387\times 10^7$  indicates the beginning of the matter dominated epoch.} \label{figbg01}
\end{figure}
At the beginning of this phase $\psi_{osc} \simeq .20$, with 
\begin{equation}
\label{A9}
V_{osc}^{1/4} \simeq2.332 \times 10^{15} \rm{GeV} \ .
\end{equation}
The Universe then undergoes a period of reheating (or inflaton domination) which ends when the bath of produced supersymmetric standard model particles comes into thermal equilibrium--the beginning of the matter-dominated era. The reheating temperature at the end of the inflation-dominated era was found to be
\begin{equation}
\label{A10}
T_{RH} \simeq 1.13 \times 10^{13} \rm{GeV} \ .
\end{equation}
An important quantity during the reheating period, see for example \cite{}, is the so-called maximal temperature defined to be
\begin{equation}
\label{A11}
T_{max}=\sqrt{V_{osc}^{1/4}T_{RH}} \ .
\end{equation}
Using \eqref{A9}, we find that
\begin{equation}
\label{A12}
T_{max} \simeq 1.623 \times 10^{14}  \rm{GeV} \ .
\end{equation}
The inflation-dominated period can be characterized by the temperature interval
\begin{equation}
\label{A13}
T_{max} > T > T_{RH} \ .
\end{equation}

When $N=1$ supersymmetry is spontaneously broken in the hidden sector, chiral matter fermions--both in the observable and hidden sectors--do not acquire soft supersymmetry breaking mass terms. However, in the oscillatory regime, the inflaton field $\psi$ develops a time-dependent VEV given by the square root of
\begin{equation}
\label{A14}
\langle \psi^{2}(t) \rangle \simeq \frac{1}{2\delta} \int_{t-\delta}^{t+\delta} \psi^{2} (\tilde{t}) d\tilde{t} \ ,
\end{equation}
where $\delta \simeq 2\pi/m$. It follows from \eqref{A2} and \eqref{A3} that, in this reheating phase, $\psi \propto H_{u}^{0}.$ Hence, observable sector chiral fermions $\psi^{\cal I}_{(o)}$ develop a time-dependent non-zero mass given by
\begin{equation}
\label{A15}
M_{\psi_{(o)}^{\cal I}}=y_{\psi \psi_{(0)}^{\cal I}\psi_{(0)}^{\cal I}} \sqrt{\langle \psi^2 \rangle}\ ,
\end{equation}
where $y_{\psi \psi_{(o)}^I\psi_{(o)}^I}$ is the Yukawa coupling parameter in the inflaton-two Weyl fermion 
$\psi_{(o)}^{\cal I}$ interaction. In contrast, the hidden sector chiral fermions cannot receive such contributions and remain massless.
Similarly, neither the observable nor hidden sector gauge bosons acquire soft supersymmetry breaking masses. However, as with the observable chiral fermions, the observable sector gauge bosons do have time-dependent masses generated by the inflaton VEV during the reheating period. 
 The  observable gauge boson masses are given by
\begin{equation}
\label{A16}
m_{W^0}=m_{W^\pm}=\frac{g_2\sqrt{\langle \psi^2\rangle}}{\sqrt{6}}\ ,\quad m_{W_R}=\frac{g_2\sqrt{\langle \psi^2\rangle}}{\sqrt{6}}\ ,
\quad m_{W_B}=\sqrt{\frac{2}{3}}g_{BL}\sqrt{\langle \psi^2\rangle}\ ,
\end{equation}
where $g_{2}$ and $g_{BL}$ are the coupling parameters for the $SU(2)_{L}$ and $U(1)_{B-L}$ gauge groups respectively.
Just as in the case of fermions, the gauge bosons in the hidden sector do not receive such contributions and, therefore, remain massless.
Finally, the case for the observable and hidden sector gauginos is more complicated, since they both receive mass contributions from soft-supersymmetry breaking. In addition, the observable sector gauginos get contributions to their masses generated by the VEV of the inflaton during the reheating period. 
The chargino and neutralino mass mixing matrix was studied in~\cite{Cai:2018ljy}. For example, the mass of the lightest 
chargino state was found to be
\begin{equation}
\label{A17}
m^2_{\lambda_{(0)}^\pm}=\frac{1}{2}\left(M_2^2+\frac{1}{3}g_2^2\langle \psi^2\rangle-\sqrt{M_2^2+\frac{2}{3}g_2^2M_2^2\langle \psi^2\rangle}   \right)\ ,
\end{equation} 
which combines the Wino gaugino  soft-supersymmetry breaking mass $M_2$ with the mass generated by the inflaton oscillation VEV. Note, however, that the hidden sector gauginos, although they do get a non-vanishing soft-supersymmetry breaking mass, do not get any further enhancement of their mass since they do not couple to $\psi$.

\section{ Supersymmetry Breaking and the Mass Spectrum}\label{sec:SUSYbreak}

In this Appendix, we analyze a possible SUSY-breaking mechanism that leads to squared scalar mass terms of the order
 of $10^{13}$GeV on the observable sector, as required by our observable sector model of inflation. 
 
 It is well known that at the non-perturbative level, \emph{gaugino condensation on the hidden sector} can induce a moduli-dependent superpotential $\hat W(S,T)$, which in turn, leads to non-vanishing F-terms $F_S$ and $F_T$ which break $N=1$ supersymmetry globally. 
 
 Let us consider our hidden sector model, which contains matter and gauge fields that transform under the gauge group $H^{(2)}=\mathcal{H}_2\times \mathcal{H}_1$. For simplicity, we will assume that the matter fields transform under $\mathcal{H}_1$ only, while 
 the gauge coupling associated with the $\mathcal{H}_2$ gauge group is such that it becomes strong at low energy, triggering the gaugino condensation mechanism, at the scale 
 \begin{equation}
 \Lambda=M_Ue^{-bS}\ ,
 \end{equation}
 where $b$ is the beta function associated with the gauge group $\mathcal{H}_2$, which must be positive for gaugino condensation to occur.
 
 To order $\kappa_4^{2/3}$, the superpotential generated by this gaugino condensate has the form
 \begin{equation}
 \hat W(S,T)=M_U^3e^{-3bS}
 \end{equation}
More generally, we could consider $\mathcal{H}_2$ to be a product of gauge groups, $\mathcal{H}_2=\sum_i \mathcal{H}_2^i$, and that each $\mathcal{H}_2^i$ has a positive beta-function $b_i$ associated, therefore allowing for separate gaugino condensates to form. Such multiple gaugino condensate systems have been studied in the context of moduli stabilization via the \emph{racetrack mechanism}. 
 Other non-perturbative contributions to the superpotential are sourced by five-brane instantons, associated with any five-branes allowed between the observable and the hidden sector in the extended orbifold geometry. Furthermore, most recent studies of the moduli stabilization problem in the heterotic vacua propose another superpotential that can fix these moduli, by turning on the flux of the non-zero mode of the antisymmetric tensor field in the bulk space. This effect generates a constant superpotential $\hat W_{\text{flux}}$ which appears at the perturbative level in the 4D effective theory, proportional to the averaged three-form flux~\cite{Cicoli:2013rwa,Ibanez:2012zz}.  The flux quantization condition~\cite{Lukas:1997rb,Gray:2007qy} constraints this constant contribution to be of the form 
\begin{equation}
\label{eq:fluxsup}
\hat W_{\text{flux}} = cM_U^3\ .
\end{equation}
 The value of the dimensionless constant $c$ is quantized such that
 \begin{equation}
  \quad c =\alpha n\ ,\quad n \in \mathbb{Z}\ ,\quad \alpha \sim \mathcal{O}(1)\ .
  \end{equation}
 
  Under all these considerations, the most general superpotential which can break $N=1$ supersymmetry in the heterotic string vacuum has the form
  \begin{equation}
  \label{eq:gen_superpotential}
  \hat W(S,T)=M_{U}^{3} \big(c+\sum_i e^{-3b_iS} \big)+\sum_x e^{n_xT}\ ,
  \end{equation}
  where the summation goes over the gauge groups $\mathcal{H}_2^i\in \mathcal{H}_2$, and over the number $x$ of five-branes. Therefore, in scenarios in which moduli are stabilized by turning on a constant flux contribution, the scale of the soft-SUSY breaking terms, as well as the masses acquired by the moduli fields, are of the order $\kappa_4^2M_U^3\approx 10^{13}$GeV, which is precisely the scale that fits our inflation model. This flux contribution can be a key tool in stabilizing the heterotic vacuum while setting the stage for high-scale SUSY-breaking.

 An F-term scalar potential  $V_{F}$ is then generated, which has the form 
given by
\begin{equation}
V_F=e^{\kappa^2_4K}\left[ g^{A\bar B}(D_A\hat W)(D_{\bar B}\hat W^*)-3\kappa_4^2{|\hat W|^2}  \right] 
\label{fly4}
\end{equation}
where
\begin{equation}
D_A\hat W=\partial_{A}\hat W+\kappa_4^2K_A\hat W\ .
\label{fly4A}
\end{equation}
The indices $A,B$ each run over the indices of \emph{all scalars} of the theory. 

The moduli stabilization problem consists in finding an explicit superpotential $\hat W(S,T)$, of the type shown in eq. \eqref{eq:gen_superpotential}, such that the F-term potential $V_F$ can be minimized with respect to the moduli fields, which in in our case are $S$ and $T$.  Explicitly, the vacuum state is defined at the fixed moduli VEVs, $\langle S\rangle$ and $\langle T\rangle$, where the first derivatives of the potential $V_F$ with respect to the moduli fields vanish,
\begin{equation}
\begin{split}
\label{eq:vac_cond1}
 \left \langle \frac{\partial V_F}{\partial S}\right\rangle= \left \langle \frac{\partial V_F}{\partial \bar S}\right\rangle= \left \langle \frac{\partial V_F}{\partial T}\right\rangle=\left \langle \frac{\partial V_F}{\partial \bar T}\right\rangle=0 \ ,
 \end{split}
\end{equation}
and all its second derivatives are positive.

In addition, one can also demand that the cosmological constant vanishes 
\begin{equation}
\Lambda=\langle V_F\rangle =0\ .
\label{eq:vac_cond2}
\end{equation}
 In this work, we will assume that it is possible to find a solution that satisfies the above conditions. Indeed, the subject of moduli stabilization in the heterotic theory is a vast one and to the knowledge of the authors, it does not have a clear solution at present. A detailed account of the moduli stabilization mechanism in heterotic vacua in which supersymmetry is broken by non-perturbative effects can be found in \cite{Cicoli:2013rwa}.

Assuming that turning on non-perturbative effects leads a stable vacuum with broken $N=1$ supersymmetry, the resultant mass spectrum of the low energy theory is the following. As discussed in \cite{Choi:1997cm,Soni:1983rm,Kaplunovsky:1993rd,Brignole:1997wnc,Martin:1997ns}:
\begin{itemize}

\item The gravitino mass 

\begin{equation}
m_{3/2}=\kappa_{4}^{2}e^{(K_{S}+K_{T})/2}|\hat W|\ .
\end{equation}
The gravitino mass is a good indicator of the scale of the masses acquired by the low-energy spectrum after supersymmetry is broken. Therefore, we define
\begin{equation}
m_{\text{SUSY}}=m_{3/2}=\kappa_{4}^{2}e^{(K_{S}+K_{T})/2}|\hat W|\ .
\end{equation}

\item Soft SUSY breaking mass terms on the observable sector. These include the  universal gaugino mass term
\begin{equation}
M_{1/2}=\frac{1}{2{\rm Re} f_1}F^{A}\partial_{A}{\rm Re}f_{1}\ ,
\end{equation}
as well as the quadratic scalar masses
\begin{equation}
m_{\cal{I} \bar{\cal{J}}}^{2}=m^{2}_{3/2}Z_{\cal{I} \bar{\cal{J}}}-F^{A}\bar{F}^{\bar{B}}R_{A \bar{B} \cal{I}\bar{\cal{J}}}\ ,
\end{equation}
where
\begin{equation}
R_{A \bar{B} \cal{I}\bar{\cal{J}}}=\partial_{A}\partial_{\bar{B}}Z_{\cal{I}\bar{\cal{J}}}-\Gamma^{\cal{N}}_{A\cal{I}}Z_{\cal{N}\bar{\cal{L}}}{\bar{\Gamma}}^{\bar{\cal{L}}}_{\bar{B}\bar{\cal{J}}}
\label{r1}
\end{equation}
and
\begin{equation}
\Gamma^{\cal{N}}_{A\cal{I}}=Z^{\cal{N}\bar{\cal{J}}}\partial_{A}Z_{\bar{\cal{J}}\cal{I}} \ .
\label{r2}
\end{equation}
In our model
\begin{equation}
Z_{\cal{I} \bar{\cal{J}}}=e^{\kappa_4^2K_T/3}\mathcal{G}_{\cal{I}\bar{\cal{J}}}\ .
\end{equation}

In general, the values of these scalar masses are of the order of the SUSY breaking scale $m_{\text{SUSY}}$, defined above. This fact becomes immediately obvious when we set $\mathcal{G}_{\cal{I}\bar{\cal{J}}}=\delta_{\cal{I}\bar{\cal{J}}}$, in which case we recover the
universal case
\begin{equation}
m_{I\bar J}=e^{\kappa_4^2K_T/3}m^2_{3/2}\delta_{I\bar J}\ .
\end{equation}

\item Hidden scalar mass terms. The hidden sector scalar fields $C^L$ also obtain soft SUSY breaking mass contributions. The formulas are identical to the scalars on the observable sector shown above; however, the $\cal{I}, \cal{J}, \dots$ indices are replaced by $L,M, \dots$.

\item Moduli mass terms. All moduli scalar masses are obtained by studying the second derivatives of the potential $V_F$, defined in eq. \eqref{fly4}, with respect to moduli fields $S$, $T$. Expanding the potential $V_F$ around the vacuum state defined above in \eqref{eq:vac_cond1} and \eqref{eq:vac_cond2}, we get
\begin{equation}
\begin{split}
V_F&=\langle V_F\rangle+\langle \frac{\partial V_F}{\partial z^A}\rangle \delta z^A+\langle \frac{\partial V_F}{\partial \bar z^{\bar A}}\rangle \delta\bar  z^{\bar A}\\
&+\langle\frac{\partial^2 V_F}{\partial z^A\partial \bar z^{\bar B}}\rangle\delta z^A\delta \bar z^{\bar B}+\langle\frac{\partial^2 V_F}{\partial z^A\partial  z^{ B}}\rangle\delta z^A\delta  z^{ B}+\langle\frac{\partial^2 V_F}{\partial \bar z^{\bar A}\partial \bar z^{\bar B}}\rangle\delta \bar z^{\bar A}\delta \bar z^{\bar B}\dots\\
&=\underbrace{\langle\frac{\partial^2 V_F}{\partial  z^{A}\partial \bar z^{\bar B}}\rangle\delta z^A\delta \bar z^{\bar B}+\langle\frac{\partial^2 V_F}{\partial z^A\partial  z^{ B}}\rangle\delta z^A\delta  z^{ B}+\langle\frac{\partial^2 V_F}{\partial \bar z^{\bar A}\partial \bar z^{\bar B}}\rangle\delta \bar z^{\bar A}\delta \bar z^{\bar B}}_{\text{mass terms}}+\dots\ .\\ 
\end{split}
\end{equation}
where $A,B=1,2$ and $(z_1,z_2)\equiv (S,T)$. We thus obtain the mass matrix which contains the masses of the scalar moduli fields. These values are always positive in a stable vacuum. Note that $(\delta z^1, \delta z^2)=(\delta S,\delta T)$ are the moduli scalar perturbations around the vacuum state,
\begin{equation}
\delta S=S-\langle S\rangle \ ,\quad \delta T=T-\langle T\rangle \ .
\end{equation}
Once the vacuum is stabilized, these scalar perturbations play the role of the new (dynamical) moduli fields of the theory.

It is more useful to express the mass matrix in terms of the real scalar and axion components of the scalar fields $z^A=\xi^A+i\eta^A$ for $A=1,2$, where $(\xi^1,\xi^2)=( s, t)$ and $(\eta^1,\eta^2)=(\sigma, \chi)$. In terms of these components, we obtain an expansion of the form
\begin{equation}
V_F=\langle\frac{\partial^2 V_F}{\partial  \xi^{A}\partial  \xi^{B}}\rangle\delta \xi^A\delta \xi^{ B}+\langle\frac{\partial^2 V_F}{\partial \eta^A\partial  \eta^{ B}}\rangle\delta \eta^A\delta\eta^{ B}+\dots\ .
\end{equation}
Note that we have assumed that in the vacuum state, $\langle\frac{\partial^2 V_F}{\partial  \xi^{A}\partial  \eta^{B}}\rangle=0$, which is generally expected. An argument for this was presented, for example, in~\cite{Dumitru:2022apw}.

In general, the mass matrices $\left[ \langle\frac{\partial^2 V_F}{\partial  \xi^{A}\partial  \xi^{B}}\rangle\right]$ and  $\left[ \langle\frac{\partial^2 V_F}{\partial  \eta^{A}\partial  \eta^{B}}\rangle\right]$ are non-diagonal. Before discussing the mass eigenstates and the associated mass eigenvalues, however, it is necessary to point out that $\frac{\partial^2 V_F}{\partial  \xi^{A}\partial  \xi^{B}}$ and $\frac{\partial^2 V_F}{\partial  \eta^{A}\partial  \eta^{B}}$ do not have mass squared units. That is because we defined our moduli states $S,T$ to be dimensionless fields, which do not have canonically normalized kinetic energy,
\begin{equation}
\begin{split}
\mathcal{L}&\supset -\kappa^{-2}_4\partial_\mu S\partial^\mu \bar S-3\kappa^{-2}_4\partial_\mu T\partial^\mu \bar T\\
&=-\kappa^{-2}_4\partial_\mu \delta s\partial^\mu \delta s-3\kappa^{-2}_4\partial_\mu \delta t\partial^\mu \delta t
-\kappa^{-2}_4\partial_\mu \delta \sigma \partial^\mu \delta \sigma-12\kappa^{-2}_4\partial_\mu \delta \chi\partial^\mu \delta \chi\ .
\end{split}
\end{equation}
Therefore, to obtain sensible mass units, it is necessary to restore the moduli fields $\delta s, \delta t, \delta \sigma, \delta \chi$ with their natural mass units. That is, the ``physical'' moduli perturbations are actually given by $\kappa_4^{-1}\delta s, \kappa_4^{-1}\delta t$, $\kappa_4^{-1}\delta \sigma$ and $\kappa_4^{-1}\delta \chi$. Those physical perturbations produce mass matrices with elements   $\kappa_4^2\frac{\partial^2 V_F}{\partial  \xi^{A}\partial  \xi^{B}}$ and $\kappa^2_4\frac{\partial^2 V_F}{\partial  \eta^{A}\partial  \eta^{B}}$, which indeed have mass squared units. The expected magnitudes of these matrix elements are also determined by the SUSY breaking scale $m_{\text{SUSY}}$. Indeed, the potential $V_F$ scales as $V_F\sim \kappa_4^2|\hat W|^2$ and therefore
\begin{equation}
\kappa_4^2\frac{\partial^2 V_F}{\partial  \xi^{A}\partial  \xi^{B}}\sim\kappa^2_4\frac{\partial^2 V_F}{\partial  \eta^{A}\partial  \eta^{B}}\sim  \kappa^{4}_4|W|^2\sim m_{\text{SUSY}}^2\ .
\end{equation}

One can then rotate the basis of physical states $(\kappa_4^{-1}\delta s, \kappa_4^{-1}\delta t)$ and $(\kappa_4^{-1}\delta \sigma, \kappa_4^{-1}\delta \chi)$ into the mass eigenstate basis  $(\delta \tilde s, \delta \tilde t)$ and $(\delta \tilde \sigma, \delta \tilde \chi)$,  thus obtaining
\begin{equation}
\begin{split}
V_F\supset m_{\tilde t}^2{{\delta \tilde t}}^2+m_{\tilde s}^2{\delta \tilde s}^2+{m^2_{\tilde \sigma}}{\delta \tilde \sigma}^2+{m^2_{\tilde \chi}}{\delta \tilde \chi}^2\ .
\end{split}
\end{equation}

Generically, the moduli mass eigenstates $\delta \tilde s$, $\delta \tilde t$, $\delta \tilde \sigma$ and $\delta \tilde \chi$ are linear combinations of the linear moduli perturbations
$\kappa_4^{-1}\delta s$, $\kappa_4^{-1}\delta t$, $\kappa_4^{-1}\delta \sigma$ and $\kappa_4^{-1} \delta\chi$ of the form
\begin{equation}
\begin{split}
\label{eq:eigen_stA}
\delta \tilde s&=x_1\kappa_4^{-1}\delta s+x_2\kappa_4^{-1}\delta t\ ,\\
\delta \tilde t&=y_1\kappa_4^{-1}\delta s+y_2\kappa_4^{-1}\delta t\ ,
\end{split}
\end{equation}
and
\begin{equation}
\begin{split}
\label{eq:eigen_st2A}
\delta \tilde \sigma&=x_1^\prime\kappa_4^{-1}\delta \sigma+x_2^\prime\kappa_4^{-1}\delta\chi\ ,\\
\delta  \tilde \chi&=y_1^\prime\kappa_4^{-1}\delta \sigma+y_2^\prime\kappa_4^{-1}\delta\chi\ ,
\end{split}
\end{equation}
where the parameters $x_1,x_2,y_1,y_2$ and $x_1^\prime,x_2^\prime,y_1^\prime,y_2^\prime$ are dimensionless.

\item Moduli fermions mass terms. Adding a non-perturbative superpotential generates new moduli fermion masses in the low-energy effective theory. These originate from the fermion bilinear terms
 \begin{equation}
 \mathcal{L}\supset
-\frac{1}{2}e^{\kappa_4^2K/2}\mathcal{D}_A{D}_B\hat W \Psi^{A}\Psi^{B}+h.c.\ 
\end{equation}
 in the supergravity Lagrangian, where $A,B$ each run over $S,T$ and
{
\begin{multline}
\mathcal{D}_AD_B\hat W=\partial_{A}\partial_{B}\hat W+\kappa_4^2(\partial_{A}\partial_{B}K\hat W+\partial_{A}KD_B\hat W+\partial_{B}KD_A\hat W)\\-\Gamma^C_{AB}D_C\hat W+\mathcal{O}(M_P^{-3})\ .
\end{multline}
}
These moduli fermion mass terms are non-vanishing in vacua in which the F-terms $F_S=D_S\hat W$ and $F_T=D_T\hat W$ are non-vanishing.

\end{itemize}

 \begin{table}[t!]
	\noindent \begin{centering}
		\begin{tabular}{c|ccc}
		\Xhline{2\arrayrulewidth}
		Sector & Field Type & Symbol & Mass \\
		\Xhline{2\arrayrulewidth}
	Moduli & scalar moduli &$\delta \tilde s,\> \delta \tilde t$ & $\sim m_{\text{SUSY}}$ \\
	&fermion moduli & $\tilde \Psi_S,\>\tilde \Psi_T$ &  $\sim m_{\text{SUSY}}$ \\
	\cline{2-4} 
	\noalign{\vskip 1mm}
	\Xhline{2\arrayrulewidth}
 Observable & matter scalars &  ${C_{(o)}^{\cal I}}$, ${\cal{I}}=1, \dots, \mathcal{N}_{(o)}$ & $\mathcal{O}(m_{\text{SUSY}})$ \\
	&matter fermions & $\Psi_{(o)}^{\cal I}$ &  $y_{\Phi \psi^{\cal I}\psi^{\cal I}}\sqrt{\langle \Phi^2 \rangle}$ \\
	&vector bosons & $A_\mu^a$, {\small$\tiny{a=SU(3),SU(2)_L,U(1)_{B-L},U(1)_{3T}}$}&  $0$ \\
	&gauginos & $\lambda^a$ &  $\mathcal{O}(m_{\text{SUSY}})$ \\
	\cline{2-4} 
	\noalign{\vskip 1mm}
	\Xhline{2\arrayrulewidth}
	Hidden & matter scalars &  ${C_{(h)}^{L}}$, ${L}=1, \dots, {N}_{(h)}$ & $\mathcal{O}(m_{\text{SUSY}})$ \\
	&matter fermions & $\Psi_{(h)}^{L}$ &  $0$ \\
	&vector bosons & $A_\mu^a$, {\small$\tiny{a=\mathcal{H}^1}$}&  $0$ \\
	&gauginos & $\lambda^a$ &  $\mathcal{O}(m_{\text{SUSY}})$ \\
	\cline{2-4} 
	\noalign{\vskip 1mm}
	\Xhline{2\arrayrulewidth}
		\end{tabular}
		\par\end{centering}
		\caption{Expected mass terms after 4D $N=1$ supersymmetry is broken \emph{softly} in the presence of one or multiple gaugino condensates on the hidden sector. The matter fermions and the gauge bosons do not obtain any mass
		contributions as a direct result of supersymmetry breaking. The rest of the fields from the observable and hidden sectors,
		obtain masses within the same magnitude range, $\mathcal{O}(m_{\text{SUSY}})$, which in our model is required to be around $10^{13}$ GeV. The masses of the moduli scalars and fermions are also proportional to $m_{\text{SUSY}}$. In general, the values moduli masses are also of the same order as $m_{\text{SUSY}}$, but it is easily possible to find cases in which they are significantly smaller, for specific non-perturbative superpotentials. }
		\label{tab:full_scan}
\end{table}

\section{Relevant Details of the $B-L$ MSSM Heterotic Vacuum}

The $B-L$ MSSM heterotic $M$-theory has been discussed in detail in the literature. It consists of Horava-Witten theory compactified to five dimensions on a CY threefold, $X$, with $h^{1,1}=3$. Here we simply present some properties of one explicit vacuum of this theory--which we use as a specific example in Section 4. Specifically, we consider the case where the hidden sector gauge group is the line bundle $L={\cal{O}}_{X}(2,1,3)$ with an anomalous $U(1)$ structure group. As shown in \cite{Ashmore:2020ocb}, this vacuum will satisfy all phenomenological and mathematical constraints if the three real components $a^{i}$, $i=1,2,3$ of the K\"ahler moduli  lie within the so-called ``magenta'' region of moduli space. This magenta region is shown pictorially in Figure 6.

\begin{figure}[t]
   \centering
\includegraphics[width=0.52\textwidth]{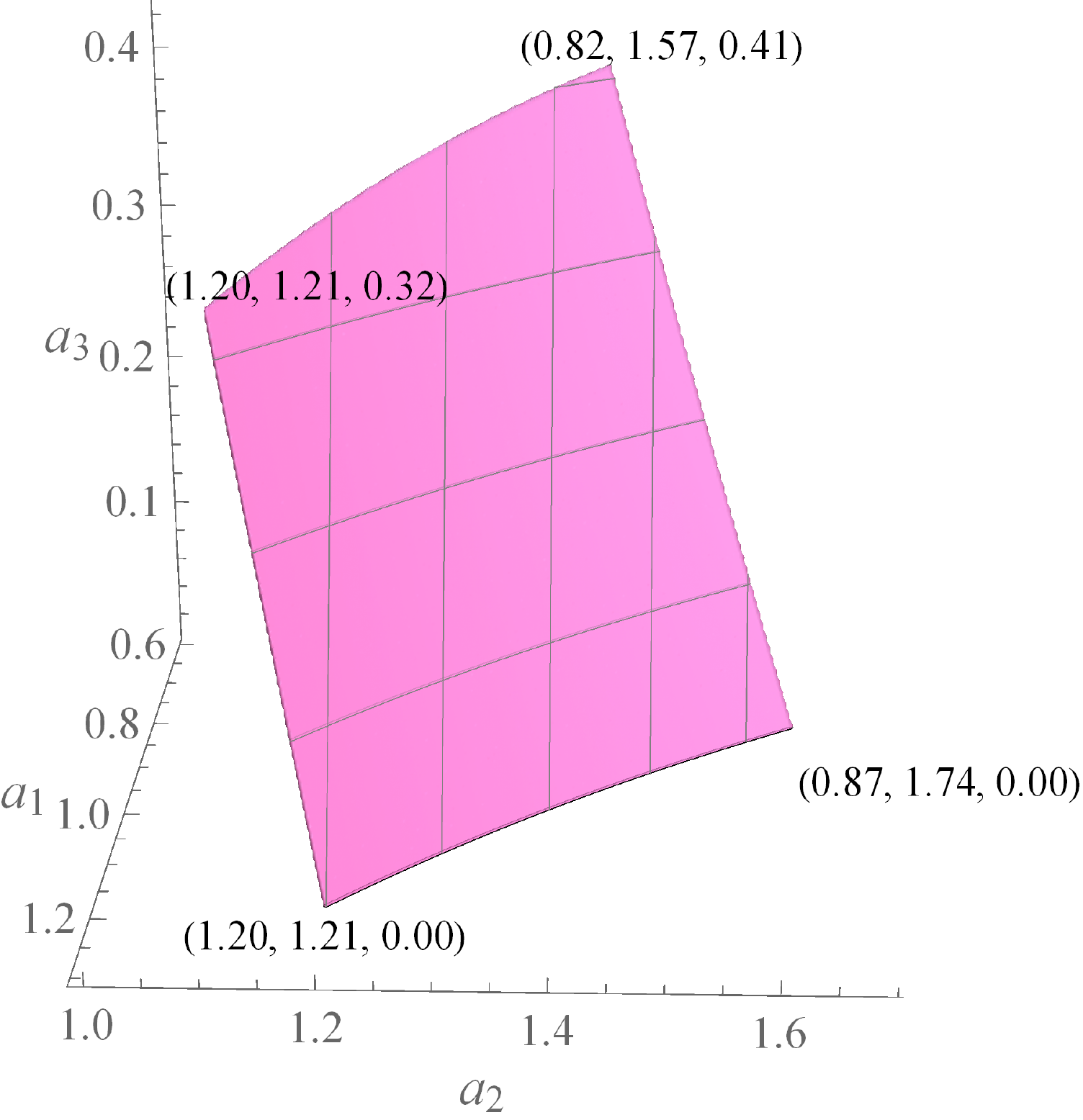}
\caption{ ``Viable'' region of K\"ahler moduli space space that satisfies all phenomenological and mathematical constraints for the line bundle $L= \mathcal{O}_X(2,1,3)$.}
\label{fig:KahlerViableRegion}
\end{figure}

For this particular line bundle, with $L \oplus L^{-1}$ embedded into $E_{8}$ so that $E_{8} \rightarrow E_{7} \times U(1)$, we showed in \cite{Ashmore:2020ocb} that the low energy hidden sector spectrum is the one reproduced in Table 2. One expects the $E_{7}$ to become strongly coupled at some mass scale which, within the context of Section 4, will be of order $10^{13}$ GeV. At this scale, all $E_{7}$ charged matter in Table 2 will condense to glueballs, gaugino condensates, and so forth. Therefore, they will not couple to moduli as presented in the text and, hence, cannot be dark matter candidates. This leaves the single anomalous $H^{*}(X,\mathcal{O}_{X})$ vector supermultiplet and the $58$ $H^{*}(X,L^{-2})$ left chiral multiplets to be considered. On the other hand, both the gauge and gaugino components of the $U(1)$ vector supermultiplet get a very large anomalous mass of ${\cal{O}}(10^{16} \rm{GeV})$ and, hence, they can be integrated out of the low energy theory. Therefore, the only possible dark matter candidates are the $58$ $H^{*}(X,L^{-2})$ left chiral supermultiplets. As discussed in Section 2 of the text, these  multiplets are of the form
\begin{equation}
\label{end1}
\tilde{C}_{(h)}^{L}=(C_{(h)}^{L}, \Psi_{(h)}^{L}), L=1,\cdots,{58}.
\end{equation}
Because all 58 chiral multiplets have identical $-1$ charges, the associated superpotential vanishes. Therefore, the supersymmetric masses of all such scalars and fermions are zero. Furthermore, the fermions cannot receive soft supersymmetry breaking masses--see Appendix B. It then follows from the analysis in Section 4 that the fermions cannot be dark matter. This then leaves only the 58 complex scalar component fields $C_{(h)}^{L}$ as dark matter candidates. In Appendix B, it was shown that these scalars can indeed get soft supersymmetry breaking masses. To the lowest order, the masses are of the form 
\begin{equation}
\label{end2}
m^{2}_{L\bar{M}}=m^{2}_{SUSY}e^{\kappa_{4}^{2}K_{T}/3} \mathcal{G}_{L \bar{M}} \ ,
\end{equation}
where $\mathcal{G}_{L \bar{M}}$ is a geometric moduli dependent hermitian matrix. The explicit moduli dependence of this matrix is unknown and, hence, one cannot, at present, compute its exactly diagonalized components.  It is conceivable that $\mathcal{G}_{L \bar{M}}=\delta_{L \bar{M}}$ . In this case all 58 scalars would have a mass of ${\cal{O}}(m_{SUSY})$. However, it is very possible that only a subset, say $\hat{N}_{(h)} < 58$, of these scalars have that mass--while the diagonal elements of $\mathcal{G}_{L \bar{M}}$  for the remaining $58-\hat{N}_{(h)}$ scalars are either much larger than unity or much smaller than unity. In that case, the masses of these $58-\hat{N}_{(h)}$ scalars are either $\gg {\cal{O}}(m_{SUSY})$ or $\ll {\cal{O}}(m_{SUSY})$ respectively. It then follows from Boltzmann suppression \eqref{bol} and \eqref{eq:OmegaDM3} that only the $\hat{N}_{(h)}$ scalars with mass approximately $m_{DM} \simeq {\cal{O}}(m_{SUSY})$ contribute appreciably to the dark matter relic density.
This then explains why the number of dark matter scalars $\hat{N}_{(h)}$ could be arbitrarily less than 58--as mentioned at the end of subsection 4.4.
\begin{table}[H]
	\noindent \begin{centering}
		\begin{tabular}{rrr}
			\toprule 
			$U(1) \times \Ex 7$ & Cohomology & Index $\chi$\tabularnewline
			\midrule
			\midrule 
			$(0,\Rep{133})$ & $H^{*}(X,\mathcal{O}_{X})$ & $0$\tabularnewline
			\midrule 
			$(0,\Rep 1)$ & $H^{*}(X,\mathcal{O}_{X})$ & $0$\tabularnewline
			\midrule 
			$(-1,\Rep{56})$ & $H^{*}(X,L)$ & $8$\tabularnewline
			\midrule 
			$(1,\Rep{56})$ & $H^{*}(X,L^{-1})$ & $-8$\tabularnewline
			\midrule 
			$(-2,\Rep 1)$ & $H^{*}(X,L^{2})$ & $58$\tabularnewline
			\midrule 
			$(2,\Rep 1)$ & $H^{*}(X,L^{-2})$ & $-58$\tabularnewline
			\bottomrule
		\end{tabular}
		\par\end{centering}
	\caption{The chiral spectrum for the hidden sector $U(1) \times E_{7}$
	with a single line bundle $L=\mathcal{O}_{X}(2,1,3)$. The Euler characteristic (or index) $\chi$ gives the difference between the number of right- and left-chiral fermionic zero-modes transforming in the given representation. We denote the line bundle dual to $L$ by $L^{-1}$ and the trivial bundle $L^{0}$ by $\mathcal{O}_{X}$.\label{tab:chiral_spectrum}}
	
\end{table}

\bibliographystyle{utphys}
\bibliography{citations2}

\end{document}